\documentclass[12pt,reprint,aps,groupedaddress,showpacs,nofootinbib,colorlinks,prd]{revtex4-1}
\usepackage{amsmath,amssymb,graphicx,amsfonts}
\usepackage{mathrsfs}
\usepackage{xcolor}
\usepackage[caption=false,justification=centering,margin={0.5cm,0cm}]{subfig}

\newcommand{\pb}{\partial \mathscr{B}}
\newcommand{\bitem}{\begin{itemize}}
\newcommand{\eitem}{\end{itemize}}
\newcommand{\be}{\begin{eqnarray}}
\newcommand{\ee}{\end{eqnarray}}
\newcommand{\bead}{\begin{eqnarray}\begin{aligned}}
\newcommand{\eead}{\end{aligned}\end{eqnarray}}

\usepackage{hyperref}
\usepackage{setspace}

\begin{document}

\title{Testing horizon topology with electromagnetic observations}

\author{Sourabh Nampalliwar}
\email{sourabh.nampalliwar@uni-tuebingen.de}
\affiliation{Theoretical Astrophysics, IAAT, University of T{\"u}bingen, Germany}

\author{Arthur G. Suvorov}
\affiliation{Theoretical Astrophysics, IAAT, University of T{\"u}bingen, Germany}

\author{Kostas D. Kokkotas}
\affiliation{Theoretical Astrophysics, IAAT, University of T{\"u}bingen, Germany}

\date{\today}

\begin{abstract}
\noindent{In general relativity without a cosmological constant, a classical theorem due to Hawking states that stationary black holes must be topologically spherical. This result is one of the several ingredients that collectively imply the uniqueness of the Kerr metric. If, however, general relativity describes gravity inexactly at high energies or over cosmological scales, Hawking's result may not apply, and black holes with non-trivial topology may be, at least mathematically, permissible. While tests involving electromagnetic and gravitational-wave data have been used to place tight constraints on various theoretical departures from a Kerr description of astrophysical black holes, relatively little attention has been paid to topological alternatives. In this paper, we derive a new exact solution in an $f(R)$ theory of gravity which admits topologically non-trivial black holes, and calculate observables like fluorescent K$\alpha$ iron-line profiles and black hole images from hypothetical astrophysical systems which house these objects, to provide a theoretical basis for new tests of black hole nature. On the basis of qualitative comparisons, we show that topologically non-trivial objects would leave a strong imprint on electromagnetic observables and can be easily distinguished from general-relativistic black holes in nearly all cases.}
\end{abstract}

\maketitle

\section{Introduction}

In general relativity (GR), the Kerr solution uniquely represents the geometry surrounding asymptotically flat, stationary black holes (BHs) in vacuum \cite{Heusler,bhuniq1}. However, motivated by a number of observational and theoretical issues (such as ultraviolet incompleteness \cite{thooft}), there is reason to believe that gravitational interaction may be modified in the strong-field or cosmological regimes \cite{clif12}. Strong-field extensions of GR especially are likely to influence the particulars of gravitational collapse \cite{harada97,cemb12,rip19}, so that astrophysical BHs, whether they be stellar-mass or super-massive, may have a non-Kerr nature (though cf. Refs. \cite{psaltis08,suv19}). As such, experiments aimed at determining the validity of the Kerr metric as a description for BHs provide a means to test GR on a fundamental level. 

One of the key underpinnings of the Kerr uniqueness result is Hawking's theorem, which states that the cross section of the event horizon of an asymptotically flat and astrophysically stable BH must have positive Euler characteristic \cite{hawk1,hawk2}. This implies that BH horizons must be topologically spherical. Hawking's result depends critically on the structure of the Einstein equations, and thus may no longer hold when introducing `correction' terms. BHs with horizons resembling arbitrary-genus Riemann surfaces are therefore mathematically permitted within certain extended theories, or even when relaxing asymptotic flatness or energy conditions; many such \emph{topological black hole} solutions have been found in GR with negative cosmological constant \cite{Vanzo,chen15,horo18} or with exotic matter \cite{peters79,gerhar82,xanth83}, and in modified theories of gravity \cite{gbtop,diltop,clifbartop}. Moreover, toroidal horizons have been found to form in numerical \cite{shap95,emp18} and analytic \cite{mann97,spivey00} studies of gravitational collapse, though their observational status is relatively unconstrained at present \cite{bambi11}.

Techniques for studying the nature of BHs can be broadly categorized as gravitational-waves based~\cite{Berti:2018cxi,Berti:2018vdi} or electromagnetic-waves based~\cite{Bambi2015,Psaltis:2018xkc}. The two techniques can be comparable~\cite{Cardenas-Avendano:2019zxd} and are often complimentary~\cite{Branchesi:2016vef,Margalit:2019dpi,Baker:2020apq}. In this work, we focus on electromagnetic-waves based techniques. In particular, we look at iron-line spectroscopy and black hole imaging. Iron-line spectroscopy refers to relativistic broadening of the Fe K$\alpha$ emission lines from accretion disks around stellar-mass BHs in X-ray binaries (e.g.~\cite{mill06,mill09}) and super-massive BHs in active galactic nuclei (e.g.~\cite{rich98,baron18}). An analysis of the broadening effect and associated ray-tracing of X-ray photons leaving the disk and reaching the detector can thus be used to test the local geometry surrounding the host BH~\cite{rey13,relxillnk,Nampalliwar2018}. BH imaging refers to the technique of combining radio wave data using very-long-baseline interferometry, resulting in an angular resolution of the size of the event horizon of a BH~\cite{Akiyama:2019cqa}. The technique has the potential to probe the spacetime close to the event horizon~\cite{bhshad2,bhshad3}. 


It is the purpose of this paper to, using an explicit exact solution in a well-motivated $f(R)$ theory, investigate astrophysical signatures that the horizon topology may imprint on the iron-line and BH imaging observables. To compute the iron-line, we use the framework of \textsc{relxill\_nk}~\cite{relxillnk,Abdikamalov:2019yrr} and implement a new set of BH solutions that allow for toroidal or spherical event horizons (the standard Schwarzschild solution of GR being a special case of the spherical topology solutions) in a system with a geometrically-thin optically-thick accretion disk. We compare the lines for the two different topologies, and for various values of a deformation parameter to compare BHs within a specific topology. We find that the horizon topology imprints a clear signature on the line profile and conclude that the iron-line observable is able to distinguish topologically non-trivial objects easily. We investigate the effect of the horizon topology on the apparent boundary~\cite{Falcke:1999pj}, finding that it is unaffected by the horizon topology. We also compute the black hole image~\cite{Vincent:2020dij} in geometrically-thick optically-thin disks and find that it is especially sensitive to the horizon topology and, except for fine tuning of the free parameters, this observable is also able to readily distinguish between the topologically non-trivial objects.

This paper is organized as follows. In Sec.~\ref{sec:hawking} we introduce the concept of a topological BH, and describe under what circumstances they may emerge within a given theory of gravity. Sec.~\ref{sec:metric} introduces a new exact solution describing a topological BH in a particular class of $f(R)$ theories of gravity, while Secs.~\ref{sec:xrs} and \ref{sec:imaging} are dedicated to iron-line spectroscopy and direct imaging tests of the aforementioned metric as a descriptor of astrophysical BHs, respectively. A discussion follows in Sec.~\ref{sec:discuss}. 

\section{Topological black holes\label{sec:hawking}}

Hawking's topology theorem, the validity of which depends on the structure of the gravitational action and the properties of any active non-gravitational fields, states that the cross section of the event horizon of a stationary BH must be topologically spherical \cite{hawk1,hawk2}. Although we will not present the details (which can be found in Chapter 9.3 of Ref. \cite{hawk2}), it is instructive to outline the proof of the theorem to show how it might break down in an extended theory of gravity. 

Assuming that a suitable generalization of the rigidity theorem holds~\cite{Racz:1999ne,joh13}, an event horizon in a stationary space-time $(\mathcal{M},\boldsymbol{g})$ is also a Killing horizon, the null-generator of which we denote by $\boldsymbol{K}$. There is another future-directed null vector orthogonal to the horizon, $\boldsymbol{n}$, which can be normalized as $n_{\mu} K^{\mu} = -1$ without loss of generality. On the horizon cross section, which we call $\pb$ as in \cite{hawk2}, a metric $\boldsymbol{\beta}$ is induced from the 4-dimensional parent with components $\beta_{\mu \nu} = g_{\mu \nu} + K_{\mu} n_{\nu} + K_{\nu} n_{\mu}$. The Gauss-Bonnet theorem and Codazzi equations now allow one to relate the topology of the BH boundary to geometric quantities on $\mathcal{M}$, viz. 
\begin{equation} \label{eq:gauss}
\chi(\pb) = \frac {1} {4 \pi} \int_{\pb} d^{2} x \sqrt{\beta}  R_{\mu \nu \sigma \rho} \beta^{\mu \sigma} \beta^{\nu \rho},
\end{equation}
where $R_{\mu \nu \alpha \beta}$ is the Riemann tensor on $\mathcal{M}$ and $\chi$ is the Euler characteristic, which is related to the genus\footnote{Not to be confused with Newton's constant which, together with the speed of light, we set to 1 throughout.} (i.e. the number of topological holes) $G$ of the surface $\pb$ through $\chi = 2\left( 1 - G \right)$ \cite{Heusler}. A positive Euler characteristic indicates a spherical topology, since $G=0$ would be the only permitted value. Making use of geometric properties of the horizon (e.g. that it has vanishing shear and expansion), manipulations of expression~\ref{eq:gauss} eventually lead to \cite{hawk1,hawk2}
\begin{equation} \label{eq:gaussbon}
\chi(\pb) > \frac {1} {2 \pi} \int_{\pb}  d^{2}x \sqrt{\beta}  K^{\mu} n^{\nu} \left( R_{\mu \nu} - \frac {1} {2} R g_{\mu \nu} \right)  ,
\end{equation}
where $R_{\mu \nu} = R^{\alpha}_{\mu \alpha \nu}$ is the Ricci tensor. Inequality~\ref{eq:gaussbon} is the essence of Hawking's result (see also Refs. \cite{gall1,gall2} for analogous results in higher dimensions). Note that we have not yet used any particular set of field equations. In GR, from the Einstein equations,
\begin{equation} \label{eq:einstein}
R_{\mu \nu} - \frac {1} {2} R g_{\mu \nu} = 8 \pi T_{\mu \nu},
\end{equation}
we see that the term within the parentheses in Eq.~\ref{eq:gaussbon} is precisely the stress-energy tensor, which, if sufficiently well behaved (i.e. abiding by the dominant energy condition \cite{Heusler}), implies that the integral is non-negative and thus that $\chi(\pb) > 0$, so that the horizon must be topologically spherical;  $\pb \cong S^{2}$. 

In the simplest modification to GR where one includes a cosmological constant, it is easy to see that the right-hand side of Eq.~\ref{eq:gaussbon} may be \emph{negative} in asymptotically anti-de Sitter (AdS) universes, thereby not excluding the possibility of topological BHs \cite{Vanzo,chen15,horo18}. If there is any matter of negative energy-density within the spacetime $\mathcal{M}$, the Euler characteristic may also be non-positive \cite{peters79,gerhar82,xanth83}. More generally, the inclusion of modified gravity terms may also imply that $\chi(\pb) \leq 0$ is permitted under certain conditions \cite{brans72,mish18}. As such, one may place constraints on the nature of strong-field gravity by identifying what signatures the Euler characteristic of a BH boundary might imprint on experimental data. 

\subsection{$f(R)$ gravity}

To provide a concrete example of a theory which permits topological BH solutions, we consider the $f(R)$ class of theories. In these theories, the Ricci scalar, $R$, in the Einstein-Hilbert action is replaced by some function of this quantity, $f(R)$. The vacuum field equations read (see e.g. \cite{felice} for a review)
\begin{equation} \label{eq:fofr1}
0 = f'(R) R_{\mu \nu} -  \frac {f(R)} {2} g_{\mu \nu} + \left( g_{\mu \nu} \square - \nabla_{\mu} \nabla_{\nu} \right) f'(R),
\end{equation}
where $\square = \nabla_{\mu} \nabla^{\mu}$ denotes the d'Alembert operator. These field equations reduce to those of GR (Eq.~\ref{eq:einstein}) in vacuo for $f(R) = R$.

We consider the class of theories governed by 
\begin{equation} \label{eq:explicittheory}
f(R) = \left(R - \Lambda \right)^{1+ \delta},
\end{equation}
where $\delta \geq 0 $ is a small parameter and $\Lambda$ is an (effective) cosmological constant. Theories of this form have a number of appealing properties, such as being scale-free (aside from $\Lambda$), possessing the correct Newtonian and weak-field limits \cite{rox77}, as well as having an obvious GR limit. In particular, as $\delta \rightarrow 0$, we recover GR with a cosmological constant. Moreover, the class of theories governed by Eq.~\ref{eq:explicittheory} are inaccessible to many perturbative techniques used to test $f(R)$ theories, because the function $f$ is not analytic for general $\delta$; this means that a Taylor expansion of the form $f = a_{0} + a_{1} R + a_{2} R^2 + \cdots$ cannot encapsulate the physics of the theory even when high-order terms are kept, and so various weak-field constraints are not immediately applicable \cite{felice}. Strong-field tests are therefore especially pertinent to theories of the form Eq.~\ref{eq:explicittheory}.

Theories defined by Eq.~\ref{eq:explicittheory} for $\Lambda = 0$ have been studied in detail by Clifton and Barrow \cite{clifbar05} (see also \cite{suvmel16}), who placed tight constraints on the parameter $\delta$. Nevertheless, as evidenced by expression~\ref{eq:gaussbon}, \emph{any} value of $\delta \geq 0$ implies that the theory may admit topological BHs for some $\Lambda$. We present some exact solutions to the field equations \ref{eq:fofr1} for the theory \ref{eq:explicittheory} in Sec.~\ref{subsec:bhsol} below.

\section{Metric structure\label{sec:metric}}

In this paper, our main interest is in testing whether or not the topological nature of a hole introduces, at least in principle, observable signatures in experimental data. We thus examine non-rotating solutions here since, although BHs are expected to rotate in reality, a consideration of static spacetimes allows for cleaner phenomenology as there are less `free' parameters to consider.

In Boyer-Lindquist $(t,r,\theta,\phi)$ coordinates, a general static, topological BH can be described by the line element \cite{mann97}
\begin{equation} \label{eq:topmet}
ds^2 = - A(r,\theta) dt^2 + \frac {dr^2} {B(r, \theta)} + r^2 d \theta^2 + r^2 H(\theta)^2 d \phi^2,
\end{equation}
where $H$ controls the topology of the surface $\pb$, and the event horizon location is defined by the vanishing of the metric potential $B$ \cite{joh13}. In general, one has three possibilities for the horizon structure depending on the genus $G$, which is encoded within the function $H$, viz.
\begin{equation} \label{eq:topology}
H(\theta) = \begin{cases} 
      \sin \theta & \text{spherical horizon; } G = 0,  \\
      \theta & \text{toroidal horizon; } G = 1, \\
      \sinh \theta & \text{higher-genus horizon; } G > 1.
   \end{cases}
   \end{equation}

\begin{figure}[!htb]
\centering
\includegraphics[width=0.493\textwidth]{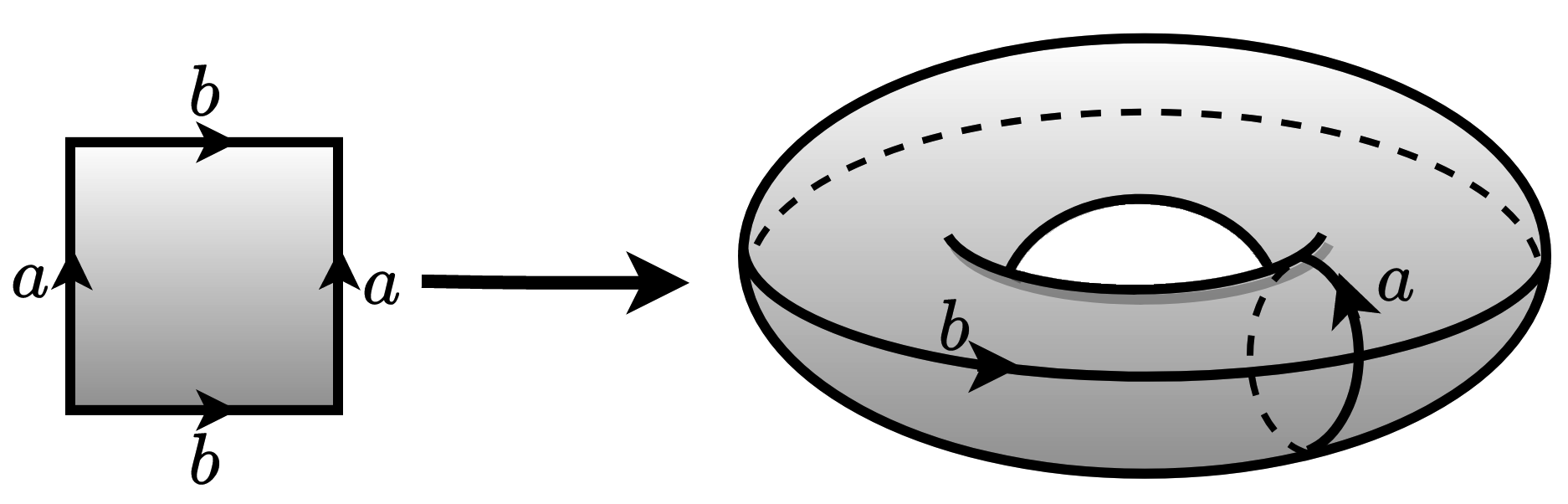}
\caption{Schematic diagram for the topological compactification of a `black plane' (left) to a toroidal BH (right), i.e., a BH whose horizon cross-section is a genus $G=1$ surface. Similar though more complicated identifications are possible in the $G>1$ case starting from the `black hyperboloid', along the lines detailed in Sec.~\ref{sec:metric}; see Refs. \cite{mann97,chen15}.  \label{fig:torus}}
\end{figure}

As written, the metric in Eq.~\ref{eq:topmet} does not contain a compact horizon except for the spherical case. However, compactification procedures can be used to remedy this. The case $H(\theta) = \theta$ prior to compactification resembles a `black plane', i.e. a naked singularity spacetime with planar symmetry. However, it is well known that the (flat) torus may be described as a quotient of the plane under certain, periodic identifications of the coordinates $(\theta, \phi)$; a square with its edges identified is topologically equivalent to a torus (see Fig.~\ref{fig:torus}), and the periodicity of $\theta$ divided by $2\pi$ defines the Teichm{\"u}ller parameter for a Riemmanian torus \cite{spivey00}. The case $H(\theta) = \sinh \theta$, on the other hand, appears as a hyperboloid. In general, under the transformation $\hat{\theta} = \tanh (\theta / 2)$, we can map a sheet of this hyperboloid to the Poincar{\'e} disk. The Killing-Hopf theorem tells us that we can identify points on this disk through some discrete subgroup of the isomorphism group $\text{SO}(2,1)$ to yield a compact surface of negative curvature with genus $G > 1$ \cite{mann98}; see also Refs. \cite{mann97,chen15} for more details.
We will not perform these identifications explicitly throughout our analysis, though it is important to note that the objects detailed herein do actually describe BHs. 

\subsection{An exact solution\label{subsec:bhsol}}

Here we present an exact, topological BH solution to the $f(R)$ theory given by Eq.~\ref{eq:explicittheory}. The simple metric defined by
\begin{equation} \label{eq:afunct}
A = 1 - \frac {2 M} {r},
\end{equation}
and
\begin{equation} \label{eq:bfunct}
B = \frac {\left( r - 2 M \right) \left[ \left( 3 \Sigma - r^3 \Lambda \right) H(\theta) - 6 r H''(\theta) \right]} {3 r \left(2 r - 3 M \right) H(\theta)},
\end{equation}
where $\Sigma$ is a constant of integration and $M$ is the BH mass, is a solution to the field equations~\ref{eq:fofr1} for the function $f$ given by Eq.~\ref{eq:explicittheory} with $\delta > 0$ for all three cases defined in Eq.~\ref{eq:topology}. To the best of our knowledge, this solution is reported here for the first time.

The $tt$-component (Eq.~\ref{eq:afunct}) of this new metric is identical to that of the Schwarzschild metric, though the $rr$-component (Eq.~\ref{eq:bfunct}) depends explicitly on the topology through $H$. Setting $H = \sin \theta$, $\Sigma = -3 M$, and $\Lambda = 0$ returns the Schwarzschild metric. The spacetime possesses an event horizon at the Schwarzschild radius $r= 2 M$, though it also contains a cosmological horizon at a value of $r$ which depends on $H$, the constant of integration $\Sigma$, and the cosmological constant $\Lambda$ (See Appendix~\ref{app:initials}). In general, the parameter $\Sigma$ can be chosen to ensure that the metric is regular everywhere between these two surfaces (see below), demonstrating that topological BHs in modified theories of gravity need not share the pathologies common to many of the known solutions in the literature \cite{Vanzo,chen15,horo18,gbtop,diltop,clifbartop}. For example, the $tt$-component of the topological de Sitter-Schwarzschild metric reads $g^{\text{TdSS}}_{tt} = -H''(\theta)/H(\theta) - 2 M /r + \Lambda r^2/3 $ (e.g. \cite{mann97}), which, in the non-spherical cases, is prone to suffer signature changes in the external region and may not contain an event horizon at all for positive mass, thereby rendering it an unsuitable candidate to describe an astrophysical compact object. 

The Kretschmann invariant $\mathcal{K} = R_{\mu \nu \alpha \beta} R^{\mu \nu \alpha \beta}$ diverges at $r=3M/2$ (in addition to at $r=0$) for $H \neq \sin \theta$, indicating the presence of a shell-like singularity which delimits the inner and outer sections of the topological BH; the horizon is located at $r=2M$ and shields the inner singularities at $r=3M/2$ and $r=0$ from an external observer. More formally, the outer horizon prevents null and timelike curves from passing from one external region to another \cite{gerhar82}.

\subsection{Some qualitative features of the new metric\label{sec:metana}}
Using the simple metric defined above, we can directly test the impact of horizon topology on observable phenomena while working within a well-motivated theory of gravity. The observables we analyze are the relativistically broadened iron line in Sec.~\ref{sec:xrs}, and the BH `image' in Sec.~\ref{sec:imaging}. Although the metric derived in the previous section is an exact solution for all three cases detailed in expression~\ref{eq:topology}, it will be most instructive to compare only the spherical and toroidal horizon cases to reduce the number of contingencies. In any case, noting that the inequality in Eq.~\ref{eq:gaussbon} is sufficient but perhaps not necessary to constrain $\chi(\pb)$, it has been argued that spherical and toroidal topologies may be the only possibilities for \emph{axisymmetric} horizon cross-sections \cite{gerhar82}. Before delving into the specifics of the relationship between observables and the spacetime geometry, we discuss some qualitative features of the metric in Eq.~\ref{eq:topmet}.

As mentioned previously, requiring that the metric is well-behaved imposes bounds on $\Sigma$. Denoting relevant variables in the spherical horizon case ($G=0$) with a subscript $S$, we have that
\be \label{eq:Bsph}
	B_S = \left ( 1-\frac{2M}{r}\right )\frac{2r + \Sigma -r^3\Lambda/3}{2r-3M},
\ee
and the Schwarzschild metric is recovered exactly for $\Sigma=-3M$ and $\Lambda = 0$. For the rest of this work, whenever we discuss the spherical horizon case, we take a vanishing cosmological constant, i.e. $\Lambda_{S}=0$, as we find that the calculated observables are largely unaffected for $|\Lambda| \lesssim 10^{-20}$. To ensure the sign of $B_S$ always remains positive outside the event horizon $r=2M$, we impose
\be
	\Sigma_S > -4 M.
\ee
In the toroidal horizon case ($G=1$), where instead a subscript $T$ is adopted on relevant quantities, the metric component $g^{rr}$ becomes
\be\label{eq:Btor}
	B_T= \left ( 1-\frac{2M}{r}\right ) \frac{\Sigma -r^3\Lambda/3}{2r-3M}.
\ee
In this case, a non-zero value of $\Lambda$ indicates that $\chi(\pb)$ from expression~\ref{eq:gaussbon} need not be positive even for $\delta \rightarrow 0$ \cite{Vanzo,chen15,horo18}. In practice, we take a sufficiently small value of $\Lambda$ ($\Lambda \sim 10^{-50}$), which ensures that the cosmological horizon is far away from both the hole and the observer, though the results are not sensitive to the exact, numerical value of $\Lambda$ provided that this latter condition is met. Demanding that $B_T$ always be positive within the region bounded by the event horizon and the cosmological horizon amounts to imposing
\be \label{eq:sigmat}
	\Sigma_T > r^3\Lambda/3,
\ee
for all radii within the domain. The lowest value of $\Sigma_{T}$ in our computations will be $0.01$, a number small enough to illustrate features of the spacetime at small $\Sigma_{T}$, and large enough to ensure that the inequality in Eq.~\ref{eq:sigmat} is satisfied. Note also that taking a negative value of $\Lambda$ and a positive $\Sigma_{T}$ ensures this automatically for all radii.

A second point to note is that, because there is no smooth transition from objects with spherical horizons to objects with toroidal horizons -- they are topologically distinct -- it is not straightforward to compare the observables in the two cases. Deviations relative to the Schwarzschild case occur in only the $g_{rr}$ component for the spherical horizon case, but in the toroidal horizon case both the $g_{rr}$ and $g_{\phi\phi}$ components change. For negligible values of $\Lambda$, expressions~\ref{eq:Bsph} and~\ref{eq:Btor} show that
\begin{equation}\label{eq:Bsmallsig}
\frac {B_S} {B_T} \sim 1 + \frac {2 r} {\Sigma}.
\end{equation}
Therefore, we expect that the observables will differ significantly between the spherical and the toroidal horizon cases for small $\Sigma$, when the ratio in Eq.~\ref{eq:Bsmallsig} is large. Moreover, within each case, the effect of $\Sigma$ will be stronger in the toroidal horizon case than in the spherical horizon case. For large $\Sigma$, despite having similar $g_{rr}$ components close to the BH, the two cases still differ in their $g_{\phi\phi}$ component. It is unclear how to anticipate whether the differences in this case would be significant or negligible, but as we will see in the next sections, the differences are indeed significant.

For the rest of this work, we pick units of mass such that $M=1$.

\section{X-ray spectroscopy\label{sec:xrs}}
X-ray reflection spectroscopy is an established technique for probing BHs in our universe. There are comprehensive reviews which describe the technique in general~\cite{Reynolds:2013qqa} and its implementation in the program to test alternative theories of gravity~\cite{Bambi2015,Abdikamalov:2019hcc,Abdikamalov:2019zfz}. We briefly describe some important concepts in X-ray spectroscopy and the reflection component below. 

\begin{figure}[!htb]
		\centering
		\includegraphics[width=0.48\textwidth]{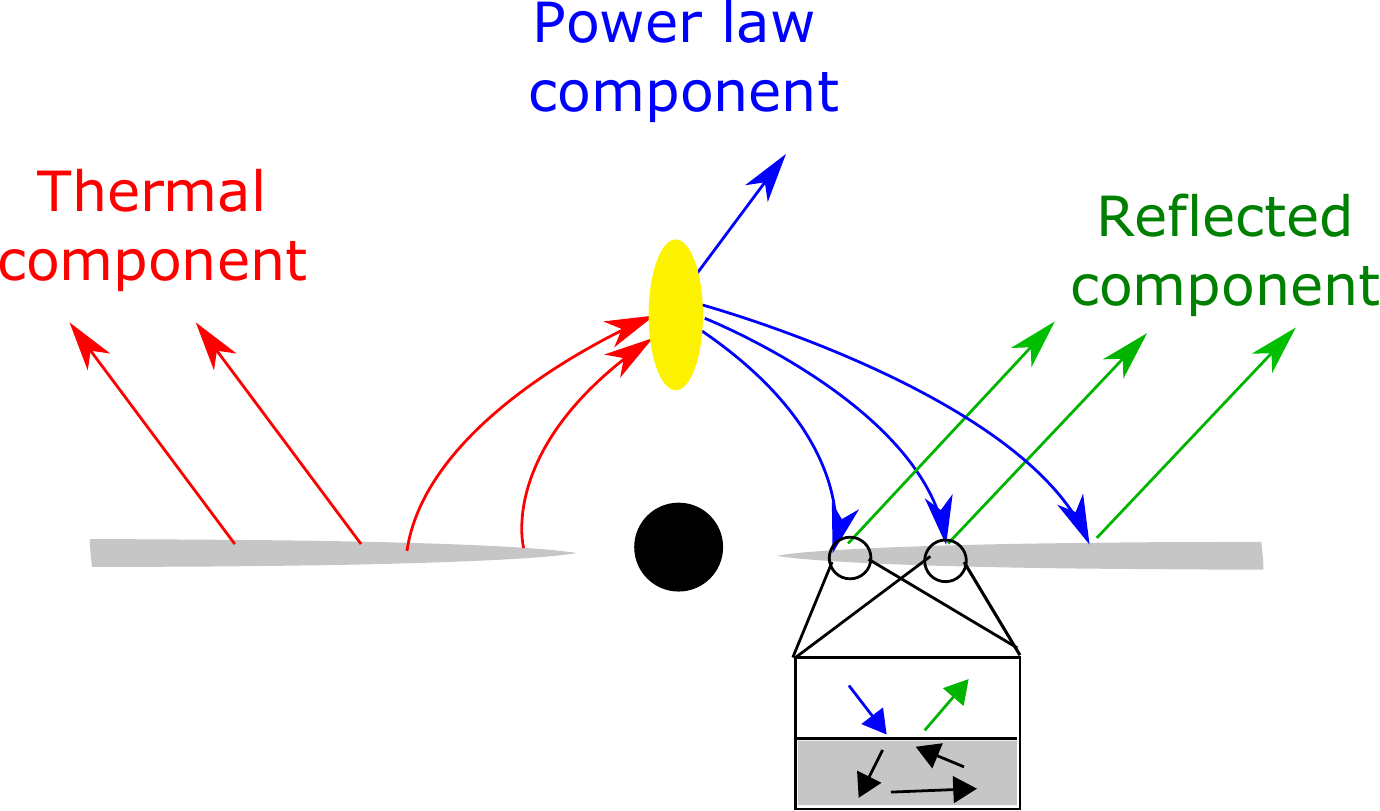}
		\caption{An illustration of the typical astrophysical system in X-ray spectroscopy. The central, filled black circle (noting again that the event horizon occurs at $r=2M$) symbolizes the BH. The geometrically-thin disk is indicated in grey, and the corona in yellow. The arrows indicate photon trajectories and are colored according to the classification shown on the figure. Adapted from~\cite{Bambi:2018thh}. See the text for more details.}
		\label{fig:diskcorona}
\end{figure}

The typical astrophysical system in the context of X-ray spectroscopy is modeled as a BH-disk-corona system, illustrated in Fig.~\ref{fig:diskcorona}. The basic components of the system are a BH, an accretion disk, and a corona. 
The accretion disk is formed out of matter that is falling onto the BH from a companion star (in the case of stellar-mass BHs) or  surrounding gas (in the case of super-massive BHs). The hot ($\sim 100$ keV) corona is a feature known primarily from observations; its morphology and composition is not clearly understood (some possibilities for what it might be include the base of an astrophysical jet or a cloud of highly energetic electrons, for instance~\cite{Bambi:2018thh}). In any case, the central BH is our primary object of interest. In the present work its geometry is described by the metric discussed in Sec.~\ref{sec:metric}. 

The total spectrum from a system like the one shown in Fig.~\ref{fig:diskcorona} can be classified in three parts, as marked in the figure. The blackbody radiation from the accretion disk forms the \textit{thermal} component. Some of this radiation gets upscattered in the corona via inverse Compton scattering and gives rise to a component with a characteristic power-law profile (i.e., the flux $\propto E^{-\alpha}$, for power-law index $\alpha$ where $E$ is the energy of the photon). Some of this \textit{power-law} component irradiates the disk, gets reprocessed inside the disk and then reflected, forming the \textit{reflection} component of the spectrum. Each component carries information about the spacetime as it traverses the gravitational well of the BH and arrives at Earth. Among the three, the reflected component is the most promising one for performing tests of gravity~\cite{Bambi:2015ldr}. A suite of data analysis tools that model the reflection component in X-rays and estimate various BH parameters can be found in Refs.~\cite{relxillnk,Abdikamalov:2019yrr}. 

\subsection{Relativistically broadened iron line\label{subsec:ironline}}	
Since our focus in this work is on the central BH, we will handle the BH neighborhood in an idealized manner aiming to capture the most prominent aspects of a realistic setup. We draw qualitative inferences from these simplified models, with the expectation that these qualitative results remain valid in more general settings. In Sec.~\ref{sec:discuss}, we discuss these simplifications and their influence on the results.  

We consider a geometrically-thin and optically-thick disk model (popularly known as the Novikov-Thorne disk~\cite{Novikov1973}), where the disk thickness is taken to be infinitesimally small. The disk is located on the equatorial plane, as anticipated for realistic disks due to the Bardeen-Petterson effect \cite{Bardeen:1975zz}. The inner edge of the disk is placed at the innermost stable circular orbit (ISCO hereafter), which for the metric under consideration is always located at $r=6$ for all $H$ choices. The outer edge of the disk does not play a big role in the calculation as long as it is at some large radius, since we use a radially decaying intensity profile. We focus our attention on Fe K$\alpha$ emission from the disk (colloquially referred to as the \textit{iron line}), because the fluorescent yield for iron is higher than for other elements. This is a typical choice made in the literature for both theoretical~\cite{Reynolds:2002np} and observational~\cite{Miller:2003wr} analyses. 

The corona is modeled in a phenomenological sense. The only role it plays in our model is to irradiate the disk, which culminates in a reflection spectrum. This latter component can be modeled by specifying the intensity profile of the disk~\cite{Wilkins2012}. We assume a simple power-law function for the intensity profile:
\be\label{eq:inten}
	I_e \propto \frac{1}{r^3},
\ee    
which is again a typical choice~\cite{Dauser:2018tzj} (it is also an accurate choice far from the inner regions of the disk if the corona is a point source close to the BH and radiates isotropically~\cite{Dauser:2013xv}). A simple extension of this function to a broken power-law (i.e., $I_e \propto \frac{1}{r^q}$ where $q$ takes different values below and above some radius) covers many more coronal geometries~\cite{Wilkins2012}, and is trivial to implement in our model. 
But in the spirit of a qualitative analysis, we contain our analysis to the simple profile.
 
The total radiation received by a observer located at a radial distance $D$ can be written as~\cite{relxillnk}
\bead\label{eq:Fobs1}
F_o (\nu_o) 
& = \int I_o(\nu_o, X, Y) d\tilde{\Omega}, 
\eead
where $I_{ o}$ is the specific intensity of the radiation at the point of detection by a distant observer, $X$ and $Y$ are the Cartesian coordinates of the image of the disk in the plane of the telescope, and $d\tilde{\Omega} = dX dY/D^2$ is the solid angle element subtended by the image of the disk on the telescope's sky. Using the Liouville's theorem, $I_{ o}$ can be related to $I_{ e}$ via the redshift factor as $I_o = g^3I_e$, where the redshift factor $g$ quantifies the change in a photon's frequency as it traverses through the spacetime from the point of emission to the point of detection. The explicit definition of $g$ and the form it takes for the accretion disk and the metric under consideration are provided in Appendix~\ref{app:disk}.
%
The flux can now be written as
\bead \label{eq:newF}
	F_o (\nu_o) = \int g^3 I_e(\nu_e, r_e, \vartheta_e) d\tilde{\Omega},
\eead 
where $\nu_e, r_e$ and $\vartheta_e$ are the photon's frequency, radial location, and emission angle relative to the normal to the disk, respectively, at the point of emission. 

For evaluating the flux, it is efficient to trace photons from the image plane of the observer back to the disk, rather than the the physically correct way from the disk to the observer. The initial conditions at the image plane are described in Appendix~\ref{app:initials}. Even with this efficient choice, in general one needs to calculate the trajectories of $\gtrsim 10^{6}$ photons. For further efficiency, we use the transfer-function formalism (first introduced in Ref.~\cite{Cunningham1975} and implemented in the context of non-Kerr metrics in Ref.~\cite{relxillnk}). This enables us to use the framework of \textsc{relxill} and \textsc{relxill\_nk}, which employ a computationally efficient scheme to integrate the transfer function to calculate relativistic reflection spectra. We use the \textsc{relline\_nk} model from the framework and modify it to compute iron lines for the spherical and toroidal metrics of this work. The details of the transfer function and its implementation in the calculation of flux are presented in Appendix~\ref{app:trf}.
\begin{figure*}[!htb]
		\centering
		\includegraphics[width=0.49\textwidth]{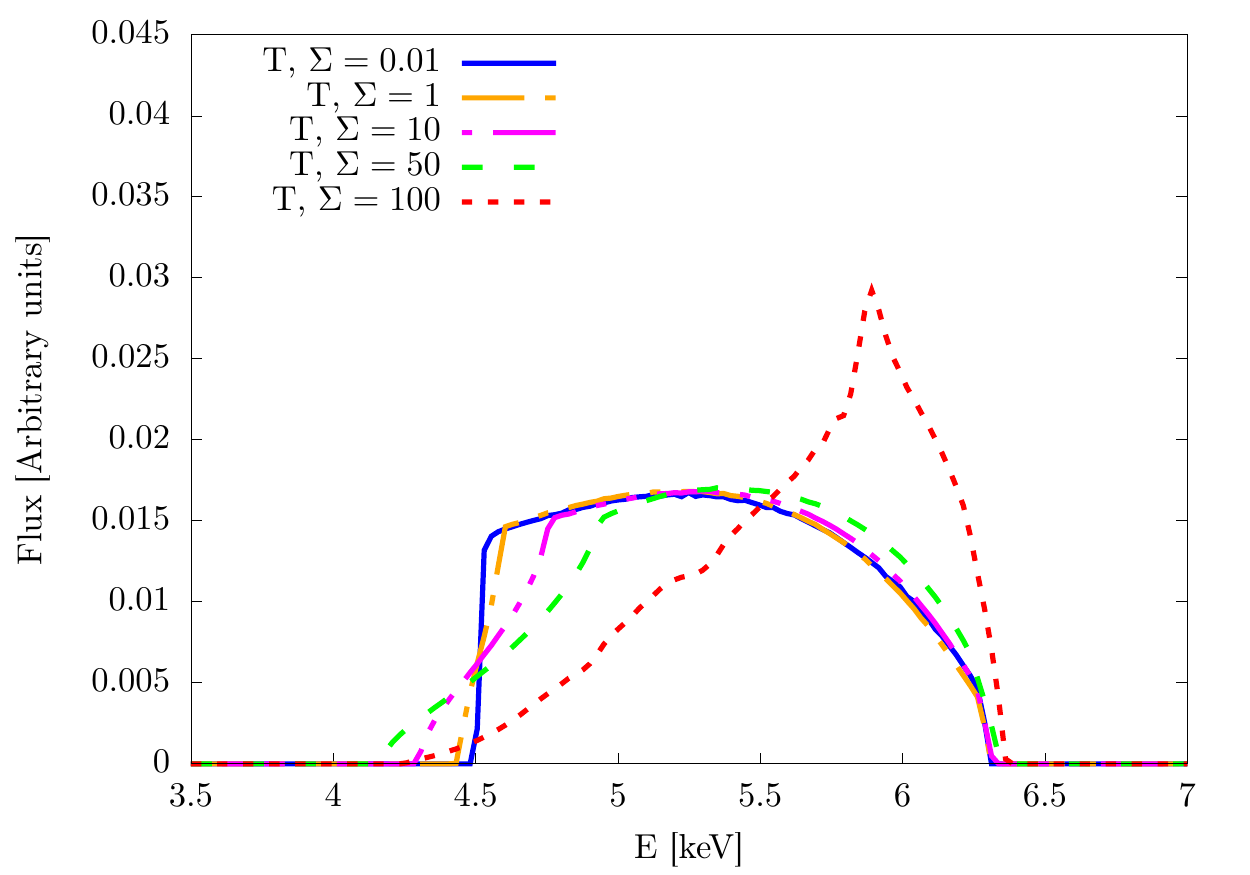}
		\includegraphics[width=0.49\textwidth]{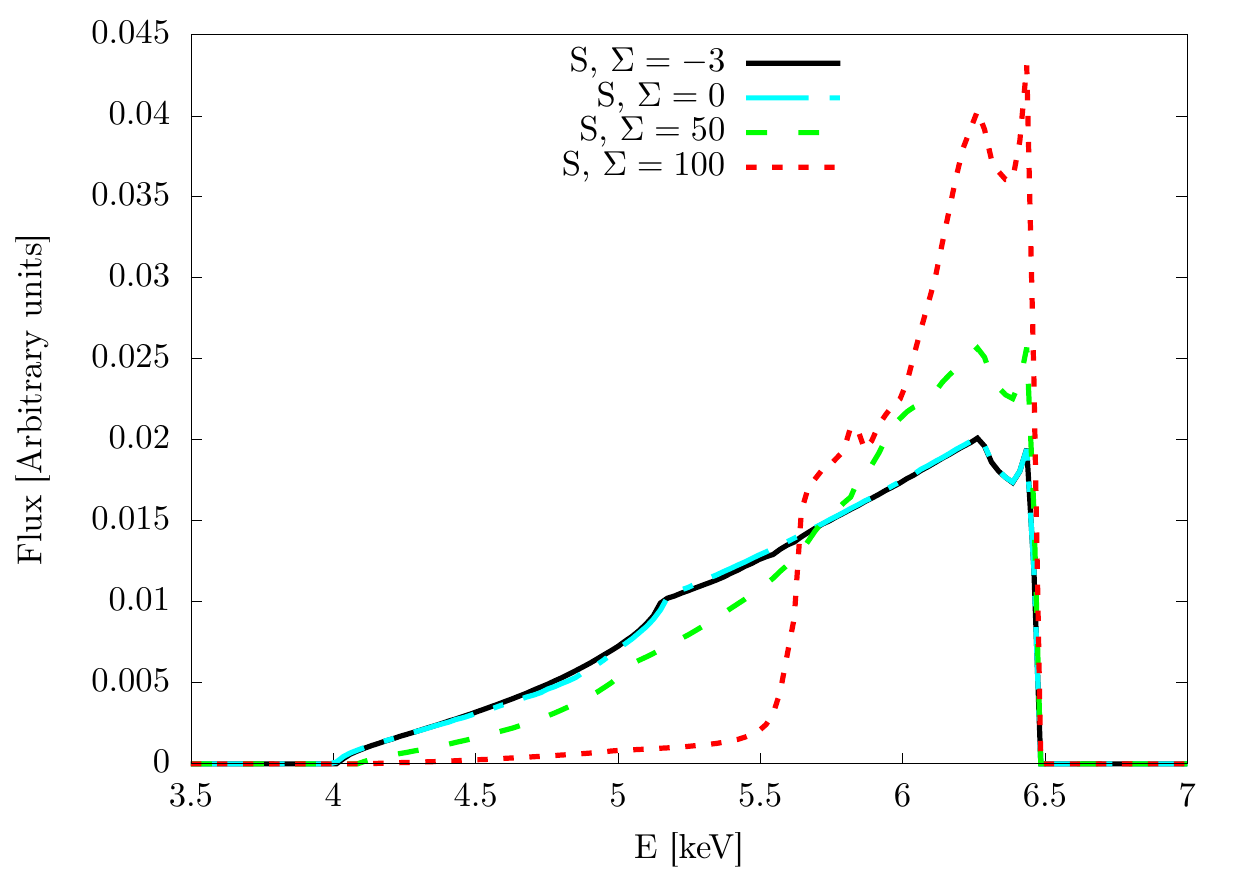}
		\includegraphics[width=0.49\textwidth]{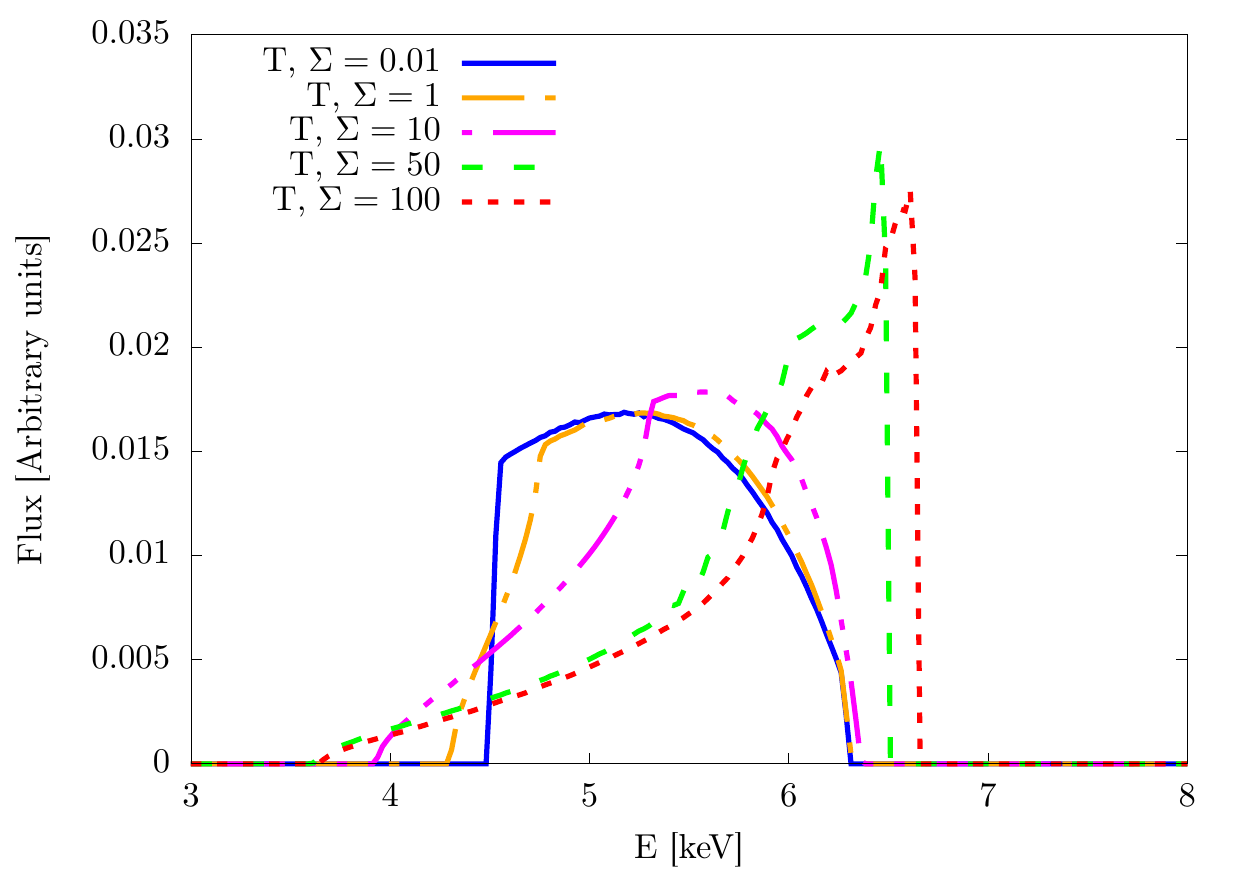}
		\includegraphics[width=0.49\textwidth]{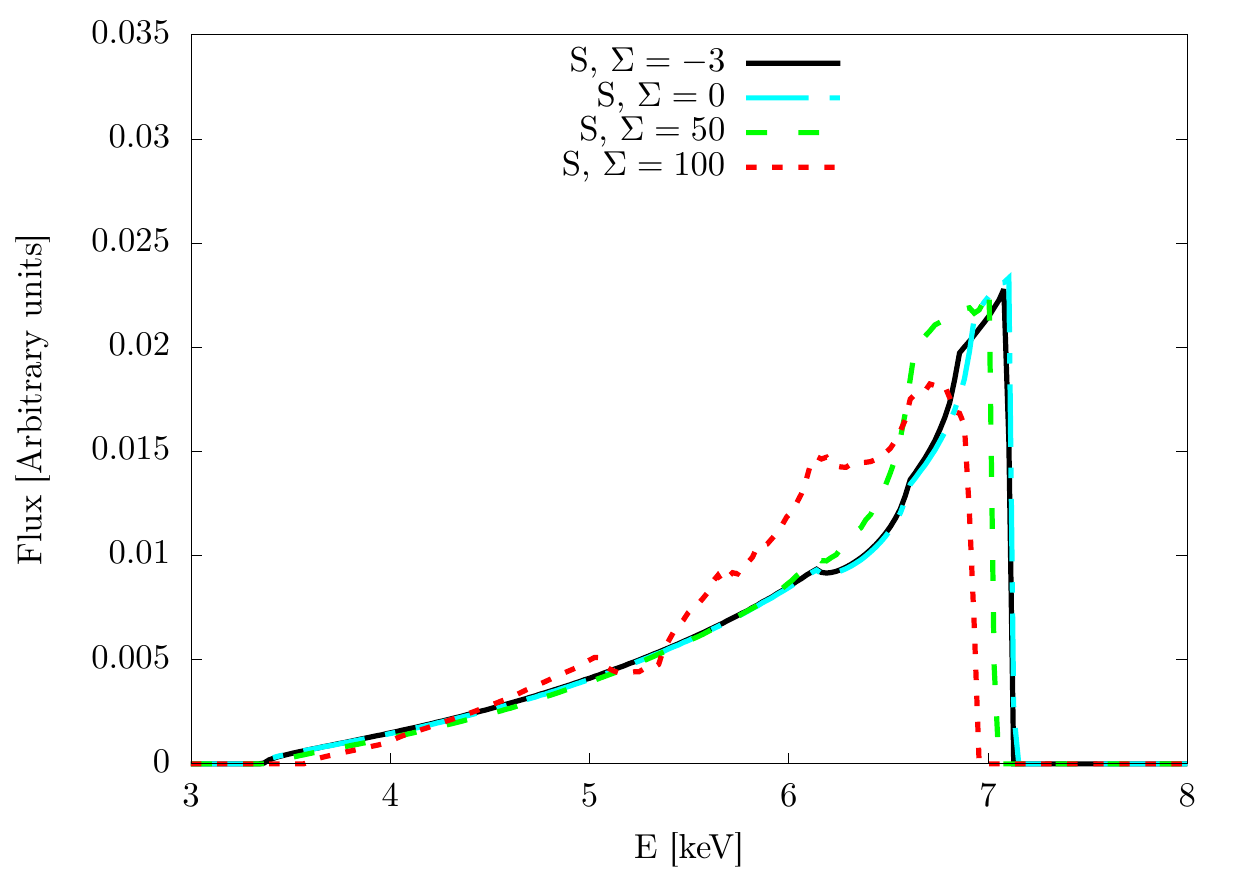}
		\includegraphics[width=0.49\textwidth]{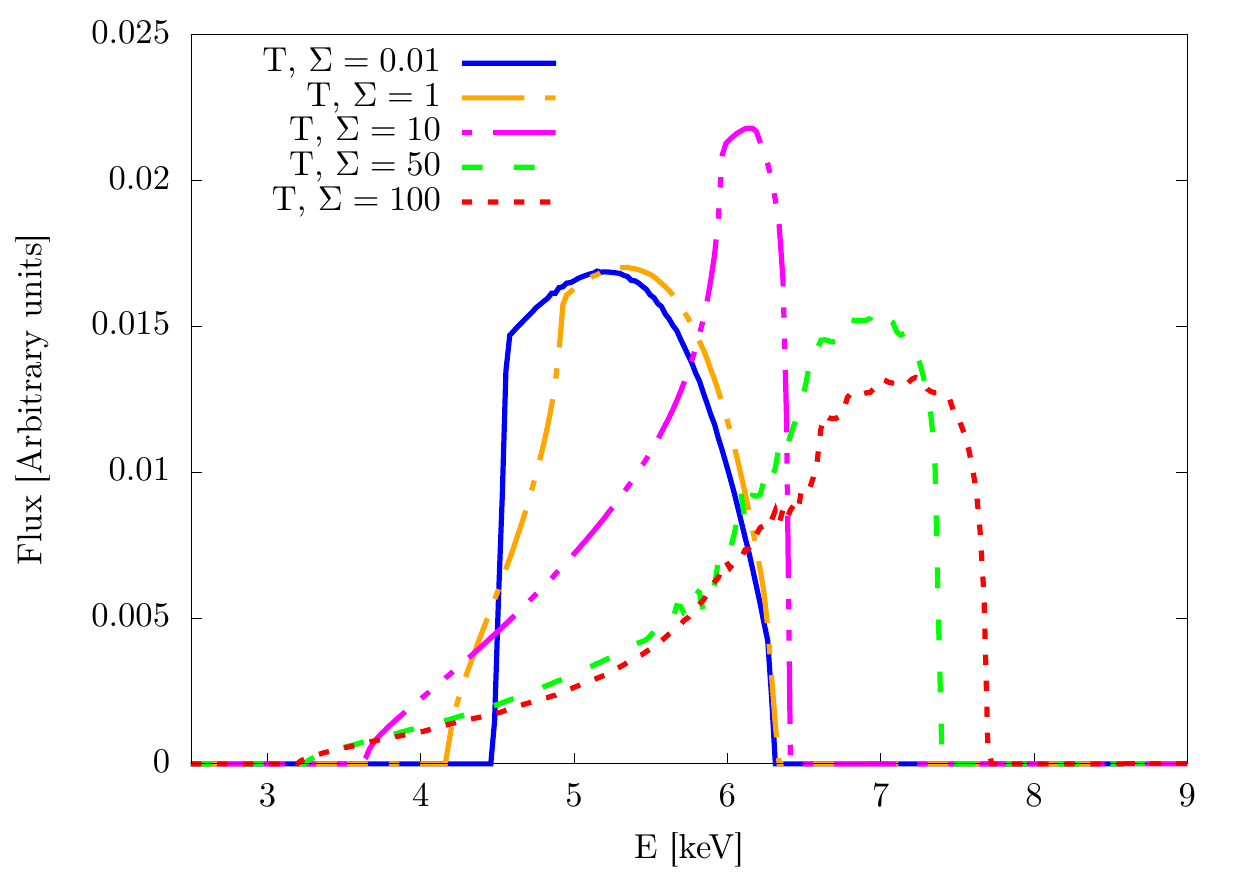}
		\includegraphics[width=0.49\textwidth]{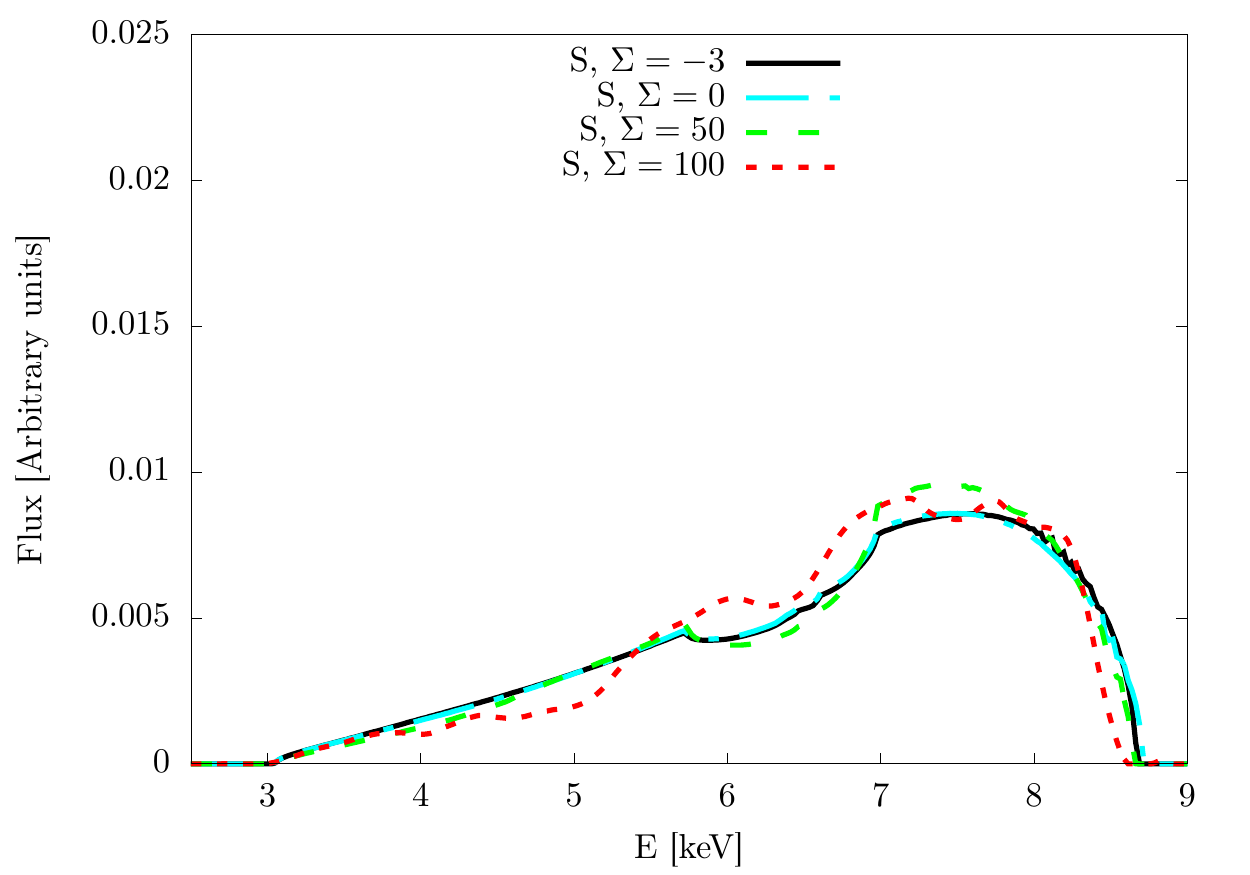}
		\caption{Relativistically broadened iron lines for several values of $\Sigma$ (labeled in each plot) in the toroidal (marked with $T$ in the labels) and spherical (marked with $S$ in the labels) horizon cases. The toroidal horizon cases are shown in the left column and the spherical horizon cases in the right column respectively. The viewing angles are $15^{\circ}$ (top row), $45^{\circ}$ (middle row) and $75^{\circ}$ (bottom row). See the text for more details.}
		\label{fig:ironlines}
\end{figure*}
\subsection{Iron-line results\label{subsec:ironlineresults}}
The iron lines calculated using the scheme described above are presented in Fig.~\ref{fig:ironlines}. The panels show lines calculated at viewing angles of $15^{\circ}$ (top row), $45^{\circ}$ (middle row) and $75^{\circ}$ (bottom row), respectively. In each row, the lines in the toroidal horizon cases are shown on the left, and those in the spherical horizon cases are shown on the right. In both cases, we plot the lines for several values of $\Sigma$. Several interesting features present themselves in these plots. 

Before discussing the role of the horizon topology and the $\Sigma$ parameter, we note some features of Schwarzschild iron-line, which serves as a benchmark. This case is represented by the black solid line in the panels on the right in Fig.~\ref{fig:ironlines}. A combination of Doppler shift, special-relativistic beaming, and gravitational redshift broadens a monochromatic line at a specific energy into a curve. The broadening depends, among other things, on the inclination of the disk relative to the observer (see Ref.~\cite{Bambi:2018thh} for a detailed discussion).

In the toroidal horizon case, for the smallest $\Sigma$ ($=0.01$), the iron lines are noticeably blue-shifted relative to the Schwarzschild line for fixed viewing angle. Among different viewing angles though, the iron lines for this particular toroidal case are nearly identical. Since typically the viewing angle affects the lines very strongly (cf. the cases in the right-hand panels), this is a novel feature of the toroidal horizon case. As $\Sigma$ increases, the lines acquire features that distinguish them from the $\Sigma=0.01$ case, but, depending on the viewing angle, even the $\Sigma=50$ case can appear quite similar to the $\Sigma=0.01$ case. For all $\Sigma$ though, the lines in the toroidal horizon case are distinct from those in the Schwarzschild case, most noticeably through the aforementioned blue-shifting and a general plateauing effect in the flux that is not observed in the latter case.

In the spherical horizon case, the  shape of the lines are largely unaffected by $\Sigma$ when it takes small values. For larger values of $\Sigma$, its influence depends on the viewing angle. For large viewing angles (when the observer's line of sight is closer to the plane of the disk) the line shape does not change significantly even for $\Sigma \lesssim 100$, whereas for a $15^{\circ}$ viewing angle, the lines are potentially distinguishable from the Schwarzschild case even at $\Sigma=50$ by the location and relative magnitudes of the peak fluxes, and the spectrum tail for $E \lesssim 5.5$ keV. 

Thus we can conclude that our first observable, the iron line, should always be able to distinguish toroidal horizon BHs from the Schwarzschild BH, and only sometimes be able to distinguish general spherical horizon BHs from the Schwarzschild BH. In the next section we will see that this is not the case with all observables.

\section{Black hole imaging\label{sec:imaging}}
BH imaging is the latest of the experimental techniques related to studies of compact objects that have become possible in recent times. The now well-known image of the center of the galaxy M87~\cite{Akiyama:2019cqa}, showing for the first time the shadow of a BH, has heralded a new era in tests of strong-field gravity. While the image itself was of low resolution, making tests of gravity at the level of other, existing techniques difficult~\cite{Zhu:2020cfn}, the approach in principle can be used for this purpose~\cite{Akiyama:2019eap}. In this section we describe how the horizon topology, modeled with the metric in Eq.~\ref{eq:topmet}, affects the apparent boundary and the BH image. 

\begin{figure}[!htb]
		\centering
		\includegraphics[width=0.45\textwidth]{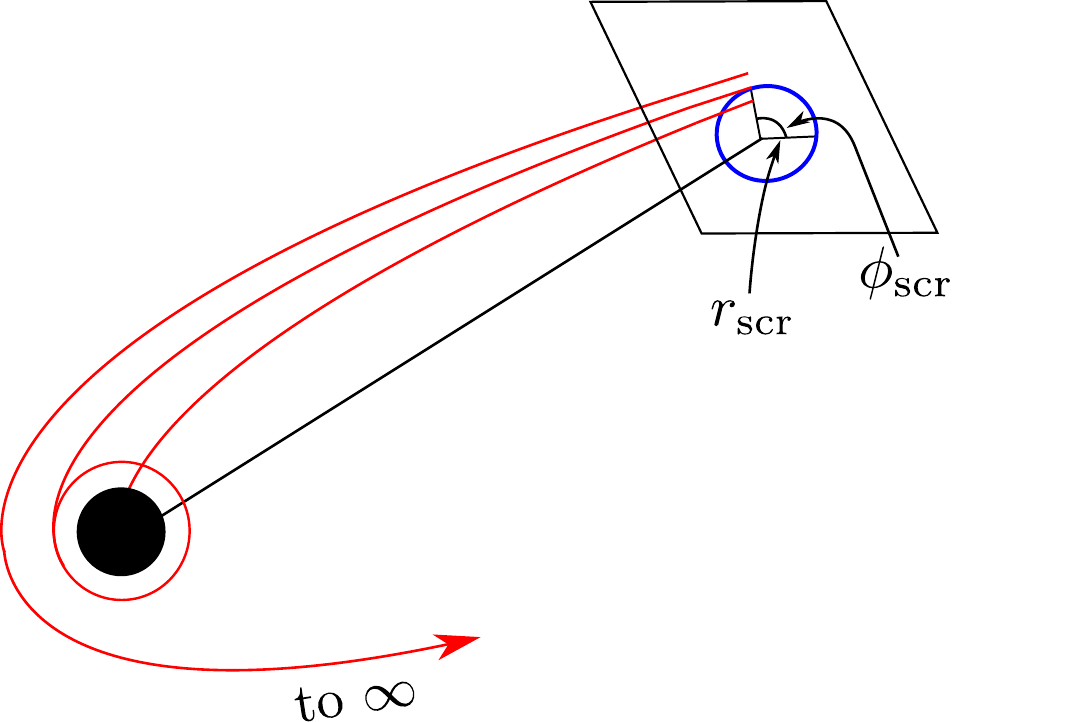}
		\caption{An illustration of the apparent boundary (colored in blue) on the image plane of the observer. Photon trajectories are shown in red, the BH by a filled, black circle and the unstable photon orbit by a red circle. The image plane is parameterized by $r_{\rm scr}$ and $\phi_{\rm scr}$.}
		\label{fig:shadow}
\end{figure}

The basic setup is shown in Fig.~\ref{fig:shadow}. As before, the observer is located at a distance $D$, far away from the event horizon (while at the same time much closer than the cosmological horizon, see Appendix~\ref{app:initials}). The photons are traced backwards in time, from the point of detection to the point of emission. The initial conditions for the photon trajectory are the same as in Sec.~\ref{sec:xrs}, and are given in Appendix~\ref{app:initials}. In this section, we introduce two new on-screen parameters, $r_{\rm scr}$ and $\phi_{\rm scr}$. These can be related to the Cartesian coordinates on the image plane $(X,Y)$ with the following simple relations:
\be
	r_{\rm scr} = \sqrt{X^2+Y^2},\\
	\phi_{\rm scr} = \tan^{-1}(Y/X).
\ee 

\subsection{Apparent boundary}
The central shadow of the super-massive BH in M87, and indeed all BHs, is closely related to the intrinsic properties of the BH. While the actual image is strongly dependent on the BH neighborhood~\cite{Akiyama:2019fyp}, which includes the accretion flow, the magnetic fields, and so on, there is one quantity that is independent of the neighborhood, and that is the so-called apparent boundary~\cite{Bardeen:1973tla}. This is defined as the boundary that divides null geodesics that fall into the event horizon and those that escape to null infinity. It can be thought of as the image of a BH illuminated by light from a distant source that is isotropically emitting and encloses the BH. For Schwarzschild BHs, this apparent boundary can be calculated analytically and occurs at a radius $r=b$, where~\cite{Luminet:1979nyg}
\be \label{eq:bparam}
	b = 3\sqrt{3}M \approx 5.2M.
\ee

\begin{figure}[!htb]
		\centering
		\includegraphics[width=0.45\textwidth]{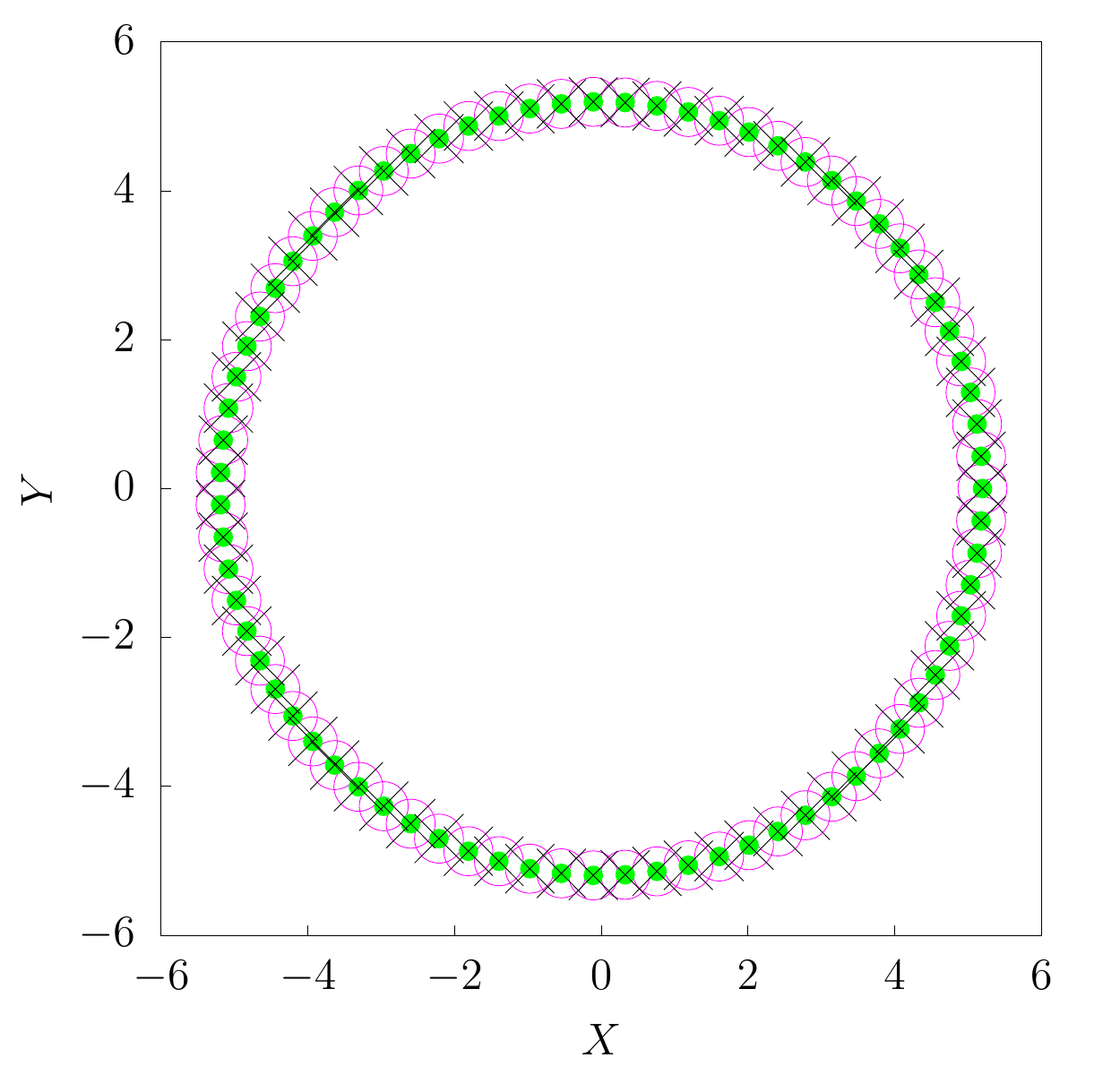}
		\caption{Apparent boundary for toroidal horizon cases with $\Sigma=0.01$ (black crosses), $\Sigma=1$ (green filled circles), and $\Sigma=100$ (magenta empty circles). See the text for more details on the construction of this image.}
		\label{fig:rings}
\end{figure}
For generic BH spacetimes an analytical expression of the apparent boundary is not possible. Rather, we must calculate it numerically. To this end, we use the following strategy. Starting with an initial guess for $r_{\rm scr}$, a photon is back-traced until it satisfies one of the two conditions: 
\bead\label{eq:condition}
 \text{I}&: \text{falls into the event horizon ($B = 0$), or}\\
 \text{II}&: \text{escapes to infinity ($r > D$),}
 \eead
where $B$ is defined in Eq.~\ref{eq:bfunct}, $r$ is the radial coordinate along the null geodesic, and $D$ is some large distance beyond which we expect the photon to continue to asymptotic infinity (set, in practice, as the radial distance of the observer from the BH). For a given $\phi_{\rm scr}$, $r_{\rm scr}$ is varied until $r_{\rm scr} - \delta r$ satisfies Condition I and $r_{\rm scr} + \delta r$ satisfies Condition II (we choose $\delta r = 10^{-6}$). This procedure is performed for 360 uniformly-spaced values of $\phi_{\rm scr}$ between $0$ and $2\pi$, which is then interpolated to determine the apparent boundary at the observer's screen. 

The apparent boundaries for three toroidal horizon cases are presented in Fig.~\ref{fig:rings}, where we take $\Sigma=0.01$ (black crosses), $\Sigma=1$ (green filled circles), and $\Sigma=100$ (magenta empty circles). We find that the apparent boundaries are all of the same size and shape within numerical precision, and match the boundary of the standard Schwarzschild case. The spherical horizon case with different values of the $\Sigma$ parameter follow the same pattern. From this, we can infer that the apparent boundary cannot distinguish either the horizon topology or the variation in the $\Sigma$ parameter from the Schwarzschild case. This is surprising at first, especially since it is in stark contrast with the iron line results of Sec.~\ref{sec:xrs} which are strongly affected by the horizon topology. This insensitivity can be understood by realizing that the outermost horizon prevents photons from illuminating the inner circle of the toroid \cite{gerhar82}; see Fig.~\ref{fig:torus}. In particular, the parameter $b$ from expression \eqref{eq:bparam} is defined as the ratio of the specific energy $E$ and angular momentum $L_{z}$ for an orbit along the photon sphere with vanishing radial momentum and acceleration \cite{chandbh22}. This quantity is thus essentially the ratio of the $t$ and $\phi$ momenta in the static case (see Appendix~\ref{app:disk}), and is independent of $H(\theta)$. Along with the fact that the $tt$-component of the metric under consideration is independent of $\Sigma$ and identical for both spherical and toroidal topologies, this implies a constant $b$ and, in turn, a constant apparent boundary. A more general case where $g_{tt}$ deviates from its Schwarzschild form would break this degeneracy and modify the apparent boundary. 

When analyzing alternative theories of gravity, it is typical to study only the apparent boundary~\cite{Amarilla:2011fx,Abdujabbarov:2016hnw,Bambi:2008jg}. 
The viability of black hole imaging for testing alternative theories is usually assessed by analyzing the effect of the parameters of the alternative theory on the apparent boundary.
In our case, although the apparent boundary is completely insensitive to the horizon topology, we shall see in the next section that a more realistic calculation of the black hole image changes the story. 

 
\subsection{Black hole image}
In reality, the actual BH image is not the same as the apparent boundary. Rather, what is observed in BH imaging is the synchrotron emission from the accretion disk around the BH~\cite{Akiyama:2019cqa}. This feature depends strongly on the astrophysics of the BH neighborhood~\cite{Pu:2018ute}: the size (how far the disk extends) and shape of the accreting matter (whether it is optically thick/thin and whether it is geometrically thick/thin), the composition and ionization of the accreting matter, and the geometry of neighboring magnetic fields all influence the character of this emission. Calculation of realistic BH images is a broad field and beyond the scope of this work, but we can get a qualitative idea by making some simplifying assumptions and calculating the BH images within these assumptions. 

\begin{figure}[!ht]
		\centering
		\includegraphics[width=0.45\textwidth]{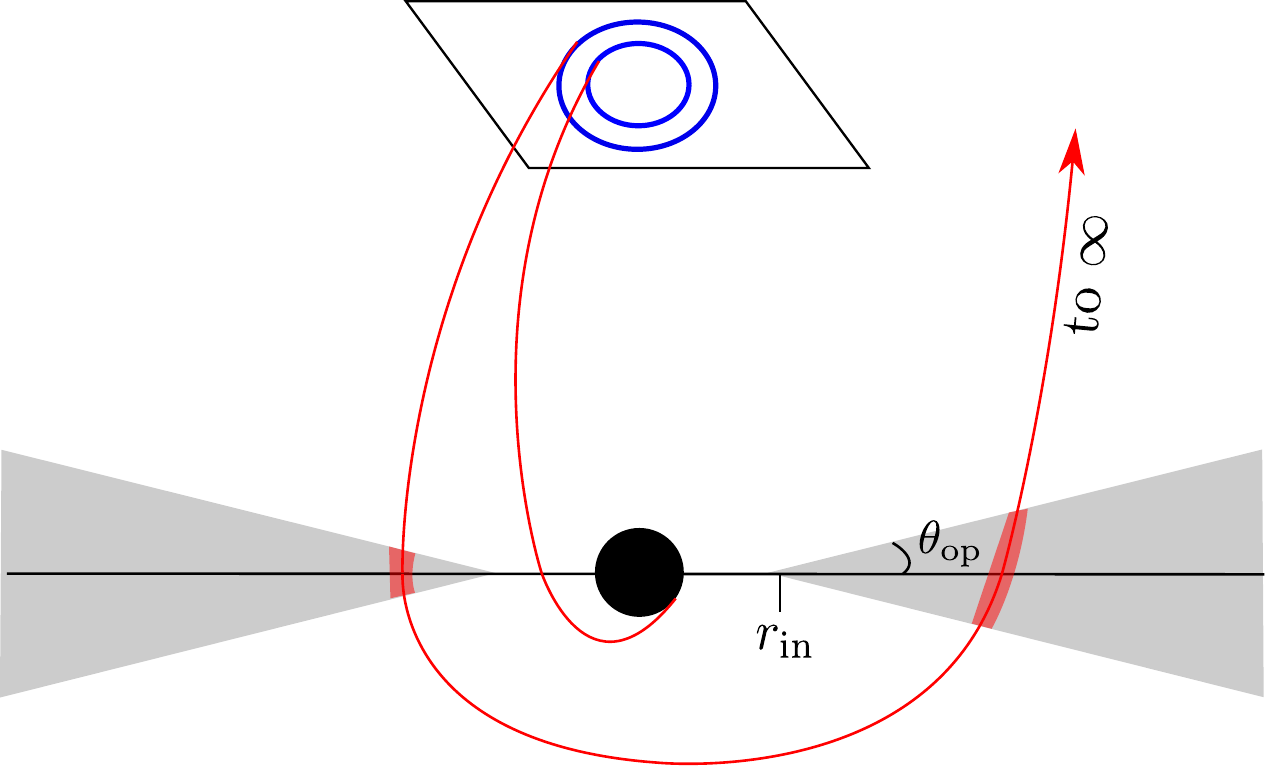}
		\caption{A schematic of the astrophysical system for modeling the BH image. The central BH is indicated with a filled black circle and the geometrically-thick disk in grey. The inner edge $r_{\rm in}$ and the opening angle $\theta_{\rm op}$ of the disk are labeled. The image plane of the observer is nearly polar in inclination, and photon trajectories are shown in red. For illustration, the part of the accretion disk through which a specific photon passes is shaded in red.}
		\label{fig:thickdisk}
\end{figure}
In contrast to the razor-thin disk of Sec.~\ref{sec:xrs}, here we assume a geometrically-thick and optically-thin accretion disk, as considered in Ref.~\cite{Vincent:2020dij} for M87. A schematic of the disk is shown in Fig.~\ref{fig:thickdisk}. The disk extends from some inner radius (which we fix at the ISCO) to some large outer radius (effectively imposed by the emissivity, see below), and has an opening angle $\theta_{\rm op}$. For simplicity, and since we are interested in only a qualitative picture, we fix $\theta_{\rm op}$ for the rest of the analysis at $30^{\circ}$. 
The material in the disk flows in circular orbits, and the velocities are independent of the height relative to the equatorial plane. Thus, the orbital velocity of particles at a given $(r,\theta)$ in the disk is set equal to the velocity of a particle in a quasi-circular geodesic on the equatorial plane at a radial location $r\sin\theta$.

Following~\cite{Falcke:1999pj,Bambi:2013nla}, we assume that the emissivity profile $j$ is monochromatic and radial, with
\be\label{eq:emis}
	j(\nu_e) \propto \frac{\nu_{\star}}{r^2},
\ee
where $\nu_{\star}$ is the rest-frame frequency. 
The total photon flux on the screen is given by (see, e.g., Refs.~\cite{Bambi:2013nla} and~\cite{Jaroszynski:1997bw})
\begin{align}
	F_{\rm obs}(X,Y) &= \int_{\gamma} I_{\rm obs}(\nu_{\rm obs},X,Y) \, d\nu_{\rm obs} \\
			&= \int_{\gamma}\int g^4 j(\nu_e) \, dl_{\rm prop} \, d\nu_e,
\end{align}
where $g$ is the redshift defined in Eq.~\ref{eq:redshifteq}, and $dl_{\rm prop}$ is the infinitesimal proper length measured in the emitter's rest frame. Using $dl_{\rm prop} = p_{\alpha}u_e^{\alpha}d\lambda$, where $\boldsymbol{u}_{e}$ is the 4-velocity according to the emitter and $\lambda$ is the path arc-length, and the expression for $g$ from Eq.~\ref{eq:redshifteq}, we get
\be
	dl_{\rm prop} = \frac{p_t}{g|p_r|} dr.
\ee 
Using this and the expression for $j(\nu_e)$ from Eq.~\ref{eq:emis}, the photon flux received by the observer can be written as
\be\label{eq:imagingflux}
	F_{\rm obs}(X,Y) = \int_{\gamma} \frac{g^3p_t}{r^2|p_r|}\,dr.
\ee
\begin{figure}[!ht]
		\subfloat[$T, \Sigma=0.01$]{\includegraphics[width=0.24\textwidth]{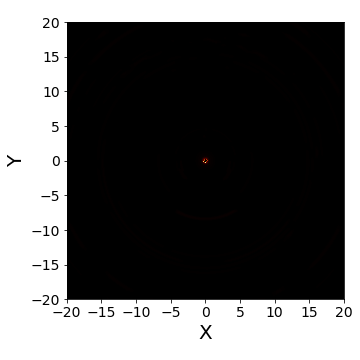}}
		\subfloat[$T, \Sigma=0.01$ Rescaled]{\includegraphics[width=0.24\textwidth]{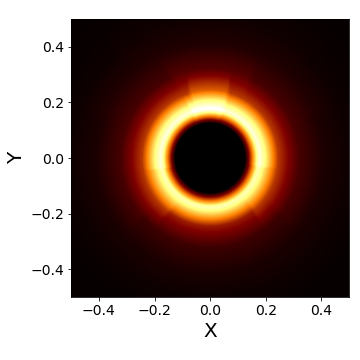}}
		
		\subfloat[$T, \Sigma=1$]{\includegraphics[width=0.24\textwidth]{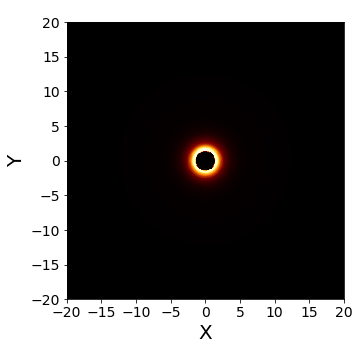}}
		\subfloat[$T, \Sigma=1$ Rescaled]{\includegraphics[width=0.24\textwidth]{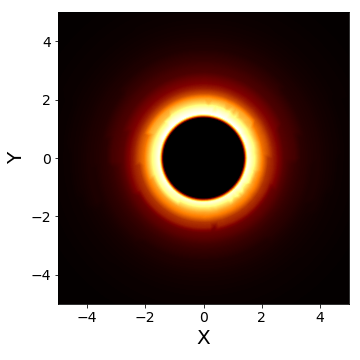}}
		
		\subfloat[$T, \Sigma=200$ ]{\includegraphics[width=0.24\textwidth]{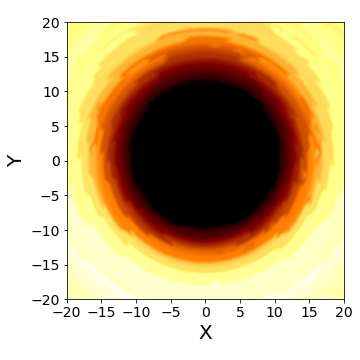}}
		\subfloat[$T, \Sigma=200$ Rescaled]{\includegraphics[width=0.24\textwidth]{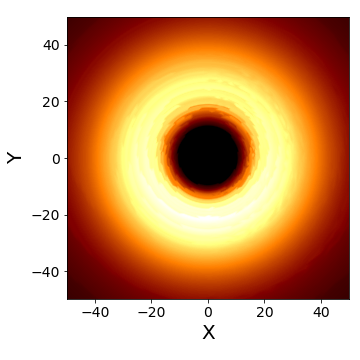}}
		\caption{\label{fig:imagestor}BH images for several values of $\Sigma$ (indicated in the subcaptions) for holes with toroidal horizons. The images in the right column show the same image as those in left column but with rescaled $X$ and $Y$ axes. See the text for further details.}
\end{figure}

\begin{figure*}[!ht]
		\centering
		\subfloat[$S, \Sigma=1$]{\includegraphics[width=0.24\textwidth]{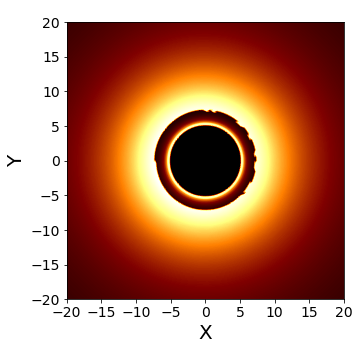}}
		\subfloat[$S, \Sigma=10$]{\includegraphics[width=0.24\textwidth]{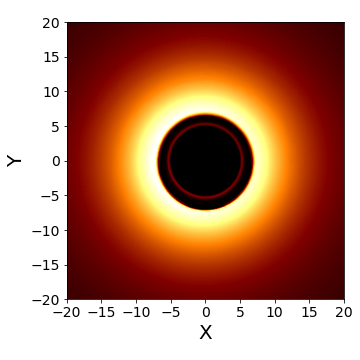}}
		\subfloat[$S, \Sigma=100$]{\includegraphics[width=0.24\textwidth]{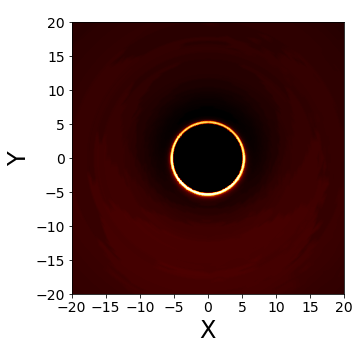}}
		\caption{\label{fig:imagessch}BH images for several values of $\Sigma$ (indicated in the subcaptions) for holes with spherical horizons.}
		
\end{figure*}
		
\begin{figure*}[!ht]
		\subfloat[$T, \Sigma=100$]{\includegraphics[width=0.24\textwidth]{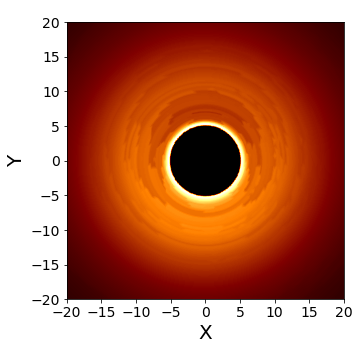}}
		\subfloat[Schwarzschild]{\includegraphics[width=0.24\textwidth]{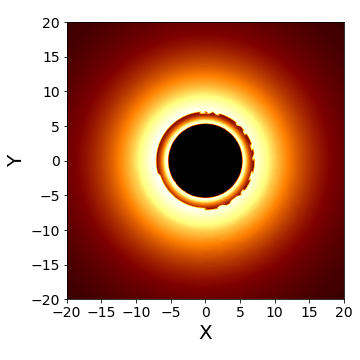}}
		\subfloat[$S, \Sigma=50$]{\includegraphics[width=0.24\textwidth]{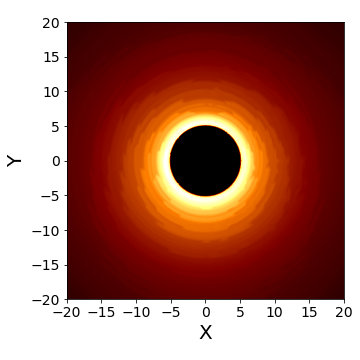}}
		\caption{\label{fig:imagesspecial}BH images for some special cases. The horizon topology and value of $\Sigma$ are indicated in each subcaption. The middle panel shows the Schwarzschild case. See the text for more details.}
\end{figure*}
To compute the image, we discretize the observer screen in terms of $r_{\rm scr}$ (150 cells from 0 to $50M$) and $\phi_{\rm scr}$ (90 cells from 0 to $2\pi$). Photons are traced backwards in time from the observer screen, as described in Appendix~\ref{app:initials}, until one of the conditions in expression~\ref{eq:condition} is satisfied. While the photon is inside the disk (the shaded region in Fig.~\ref{fig:thickdisk} represents the part of the trajectory inside the disk in one specific case), the integral in Eq.~\ref{eq:imagingflux} is calculated. This procedure is repeated for all values of $r_{\rm scr}$ and $\phi_{\rm scr}$ and the final flux-map interpolated to give a smooth image. 

\subsection{Image results}
The results of the procedure described in the previous section are shown in Fig.~\ref{fig:imagestor},~\ref{fig:imagessch} and~\ref{fig:imagesspecial}. In all cases, the viewing angle is fixed at nearly $0^{\circ}$ (i.e., the observer is nearly perpendicular to the equatorial plane, see Fig.~\ref{fig:thickdisk}). This choice is motivated by a similar orientation of the M$87^*$ system relative to Earth~\cite{Akiyama:2019fyp}. To establish a benchmark against which we will compare all the other cases, we first discuss the image of a Schwarzschild BH in this setup. This is shown in the middle panel of Fig.~\ref{fig:imagesspecial}. In this case, we see all the characteristic features associated with BH images. In particular, there is a central dark region from where no photons arrive. This is bounded by a bright ring, then a region of lower brightness, followed by a bright band that gradually fades as we go to larger radii. Although there is no direct comparison possible, we can see it as a combination of the images presented in Refs.~\cite{Vincent:2020dij} (whose disk structure we follow) and~\cite{Falcke:1999pj,Bambi:2013nla} (whose emissivity profile we follow). 

Toroidal horizon cases are presented in Fig.~\ref{fig:imagestor}. In the panels in the left column, the field of view is fixed and $\Sigma$ varies from $0.01$ in the top panel to $200$ in the bottom panel. The most striking feature of these images is the very high sensitivity of the image size to the value of $\Sigma$. From top to bottom, the central dark region goes from being practically invisible to very large, relative to the Schwarzschild case. Such a large difference in the image size would easily be detected with current observational techniques. It is important to emphasize here that the image size is not always known beforehand, so a benchmark may not be readily available. The two most important parameters that determine the size of the image within GR are the mass of the BH and its distance from us (see, for instance,~\cite{Psaltis:2008bb}). Typically these parameters are known from independent techniques (e.g., observing the motion of stars and gas in the disk around the BH~\cite{Ghez:2008ms,Gillessen:2008qv}, radiation from the infalling gas~\cite{Walsh:2013uua,Akiyama:2019eap}) but with some uncertainty. Thus the expected image size too can be estimated, but with some uncertainty. In light of this, we look at a separate feature of the images: the brightness profile. To focus on this feature, we rescale the field of view so that the volume of the central dark region in all cases is roughly equal. 
This is shown in the panels in the right column. With this normalization, we see that the $\Sigma=0.01$ and $\Sigma=1$ cases are actually rather similar. The $\Sigma=200$ case in the bottom right panel, on the other hand, is still obviously distinct from the other two. All cases are clearly distinguishable from our benchmark Schwarzschild case (middle panel of Fig.~\ref{fig:imagesspecial}) regardless.

The spherical horizon cases are presented in Fig.~\ref{fig:imagessch}. Here we find that for small values of $\Sigma$ ($\lesssim 1$), the image is nearly identical to the Schwarzschild case. This is not so surprising since, as discussed in Sec~\ref{sec:metana}, the metric in this case is mathematically similar to the Schwarzschild case (see Eq.~\ref{eq:Bsph}). For larger $\Sigma$ some distinctive features emerge, in particular for $\Sigma=10$ we see that the inner bright ring gets thinner, and the two bright bands merge (see, for instance, the right panel in Fig.~\ref{fig:imagesspecial}). For $\Sigma \gtrsim 100$, as shown in the right panel in Fig.~\ref{fig:imagessch}, the flux is completely confined near the apparent boundary (see Fig.~\ref{fig:rings}) and the image looks significantly different from both the benchmark and other cases.

Since the toroidal and the spherical horizon cases have their own characteristics and vary quite a lot in their appearance, a question arises as to whether they are always significantly different from each other. To address this question, we present Fig.~\ref{fig:imagesspecial}. The left panel shows a toroidal-horizon case, the right panel a spherical-horizon case, and the middle panel shows the image in the Schwarzschild case. We can see that the left and right images, while being significantly different from the middle image, are comparable to each other in terms of the local dispersion and flux gradients. At least with current observational techniques, it is not clear if the two could be distinguished in principle. 

We therefore conclude that our third observable, the BH image, may readily distinguish toroidal horizon BHs from Schwarzschild BHs. More general spherical horizon BHs are also distinguishable but only for large (i.e., non-Schwarzschild) values of $\Sigma$. Finally, in special cases where $\Sigma$ is fine-tuned, the images in the toroidal and spherical horizon case can look similar.
\section{Discussion\label{sec:discuss}}

A generic prediction of theories of gravity which extend GR in the strong-field regime is the emergence of non-Kerr features in BHs. These features can manifest in a number of ways, though one interesting possibility which has received relatively little attention is in the horizon topology \cite{spivey00,bambi11}: Hawking's theorem stating that BH boundaries must be topologically spherical \cite{hawk1,hawk2} need not hold in an extended theory theory of gravity \cite{brans72,mish18}, as can be seen from the inequality in Eq.~\ref{eq:gaussbon}. In this paper, we provide the theoretical basis for testing horizon topology using X-ray reflection spectroscopy (Sec.~\ref{sec:xrs}) and imaging techniques (Sec.~\ref{sec:imaging}). Using a new exact solution in the $f(R)$ class of theories, we find that a toroidal horizon compact object amplifies the low-energy part (`red region') and suppresses the high-energy part (`blue region') of the 
spectrum of K$\alpha$ iron line emissions (see Fig.~\ref{fig:ironlines}) and is, except for extreme parameter choices (unlikely to be realized for an astrophysical source), easily distinguishable from the typical BHs of GR. BH images (i.e., synchrotron radiation fluxes) also seem to display features that are unique to each topology (see Fig.~\ref{fig:imagesspecial}). On the other hand, the apparent boundary of the hole (Fig.~\ref{fig:rings}) often cannot be used to distinguish between regular and topological BHs, because an external observer cannot discern features between any inner and outer horizons by definition.

These results, though not conclusive, suggest that (static) topological BHs are unlikely to exist in reality. We note however that further tests (such as those coming from ringdown analysis \cite{scalarfield}) are needed to be more definitive, since there may be topological BHs in other theories of gravity which behave more subtly. Numerical simulations of gravitational collapse \cite{shap95,emp18} have found that toroidal holes may even exist in GR for short times post-collapse. The emergence of such transient features do not violate Hawking's theorem, which concerns stable objects, though are difficult to test with electromagnetic techniques which require long observation times. Gravitational-wave based tests would be especially helpful in this direction. Further studies of the $G>1$ case would also be worthwhile using the methods presented here. More general metric solutions where the function $H$ is more interlaced in the metric coefficients (so that $b$ depends on $H$) may change some of the effects described here, in particular the apparent boundary may become sensitive to the horizon topology. An extension of our analysis to rotating solutions, using, for instance, the Newman-Janis algorithm \cite{erbin}, would also be worthwhile in this direction. This is especially interesting since the performance of different observables may change with the introduction of spin. In particular, the iron line for some toroidal horizon cases may become degenerate with the line for the Kerr BHs, since spin amplifies the low-energy part and suppresses the high-energy part of the line, much like a toridal horizon does. This may reduce the ability of the iron-line observable to differentiate horizon topologies. The effect of the horizon topology on the black hole image, especially the huge changes in the image size (See Fig.~\ref{fig:imagestor}), is however unlikely to be mimicked by Kerr BHs~\cite{Falcke:1999pj}. The apparent boundary is also affected by the BH spin~\cite{Chandrasekhar:1985kt}, so it would be interesting to see if the degeneracy we observed here among apparent boundaries is broken in the presence of spin.

As this work is meant to be qualitative in nature, we have intentionally refrained from going into the details of two aspects. Firstly, the actual observable in X-ray spectroscopy consists of various components, as described in Sec.~\ref{sec:xrs}, and the iron line is only one part of the full spectrum. Any realistic analysis with X-ray spectroscopy will necessarily involve calculating the full spectrum. Even the system morphology, e.g., the assumptions of a geometrically-thin and optically-thick accretion disk \cite{Novikov1973} or the universality of the Bardeen-Petterson effect \cite{Bardeen:1975zz}, may not be valid for toroidal BHs. Similarly, in BH imaging, the actual disk requires relativistic magnetohydrodynamic modeling, and the radiation profile is more complex than a monochromatic power-law. The detection and analysis is also much more involved than simply recording all photons received on an image plane. Secondly, both X-ray spectroscopy and BH imaging models have several parameters apart from those we analyzed above, and some of them could be strongly degenerate with our parameters. In particular, the X-ray spectrum and the BH image depend strongly on the location of the disk inner edge, the BH spin, the emissivity index, and so on. Additionally the BH image depends very strongly on the BH mass and distance. Thus it is possible that some cases that we inferred to be distinguishable may become degenerate by varying these additional parameters. Nevertheless, we are fairly confident that the generic inferences made here, namely that horizon topology imprints a unique signature on the observables associated with X-ray spectra and BH images, will remain true.
\section*{Acknowledgments}
We thank Cosimo Bambi for useful discussions about calculating the black hole images. This work was supported by the Alexander von Humboldt Foundation. 

\appendix
\section{Initial conditions for photon trajectories\label{app:initials}}
In our ray-tracing scheme, photons are fired backwards in time, from the image plane towards the BH. Since we are dealing with a metric that is not necessarily asymptotically flat (see Eq.~\ref{eq:Btor}), writing the initial conditions for the trajectory of the photon is not straightforward. The first step is to specify the photon's initial position. As described in Sec.~\ref{subsec:bhsol}, our spacetime is endowed with both a standard event horizon (at $r_H=2M$) and a cosmological horizon. The latter is located where, following Eq.~\ref{eq:Btor}, we have
\be
	3 \Sigma - r_{\rm CH}^3 \Lambda = 0,
\ee 
or
\be
	r_{\rm CH} = \sqrt[3]{\frac{3 \Sigma}{\Lambda}},
\ee
where $r_{\rm CH}$ is the radial coordinate at the cosmological horizon. We place the observer at some distance $D$ such that
\be
	r_H \ll D \ll r_{\rm CH}.
\ee
As $D$ is also much larger than the size of the image plane, we can safely set the initial positions of photons to be equal to the observer's coordinates. Thus, at $t=0$, the photon's position vector is simply $(D,\iota,\phi)$, where $\iota$ is the inclination of the observer relative to the normal to the accretion disk, and $\phi$ is the azimuthal angle which, using the freedom provided by the axisymmetry of the spacetime, we set equal to 0. 

To specify the photon's initial 4-momentum, we use the coordinates on the image plane $(r_{\rm scr},\phi_{\rm scr})$, shown in Fig.~\ref{fig:shadow}. By setting $\phi_{\rm scr}=0$ along the direction parallel to the equatorial plane, we can relate the initial photon momentum to these coordinates as
\bead
	r_{\rm scr}\cos \phi_{\rm scr} &= \lim_{r\rightarrow\infty}\frac{-rp^{(\phi)}}{p^{(t)}}, \\
	r_{\rm scr}\sin \phi_{\rm scr} &= \lim_{r\rightarrow\infty}\frac{-rp^{(\theta)}}{p^{(t)}}.
\eead
Here, $p^{(\mu)}$ represents the photon's four momentum in the locally non-rotating frame of reference, which can be related to the global coordinates (in which the metric is presented) using orthonormal tetrads defined as~\cite{Bardeen:1972fi}
\bead
	\boldsymbol{e}_{(t)} = \frac{1}{\sqrt{g_{tt}}}\frac{\partial}{\partial t}, \\
	\boldsymbol{e}_{(r)} = \frac{1}{\sqrt{g_{rr}}}\frac{\partial}{\partial r}, \\
	\boldsymbol{e}_{(\theta)} = \frac{1}{\sqrt{g_{\theta\theta}}}\frac{\partial}{\partial \theta}, \\
	\boldsymbol{e}_{(\phi)} = \frac{1}{\sqrt{g_{\phi\phi}}}\frac{\partial}{\partial \phi},
\eead
for a spherically symmetric metric. We use these relations to get the $(p^r, p^{\theta}, p^{\phi})$ components of the initial four momentum (we set $p^{(r)}=1$). For $p^t$, we use the fact that along null geodesics, $p_{\mu}p^{\mu}=0$. So
\be
	p^t_0 = \sqrt{g_{rr}(p^r_0)^2 + g_{\theta\theta}(p^{\theta}_0)^2 + g_{\phi\phi}(p^{\phi}_0)^2}
\ee
where the subscript $0$ implies $t=0$.

\section{Photon trajectories and the accretion disk\label{app:disk}}
To calculate the trajectory of the photons and the gas in the disk, we use the standard geodesic equations of GR:
\be
	\frac{d^2x^{\mu}}{d\lambda^2} = \Gamma^{\mu}_{\alpha\beta}\frac{dx^{\alpha}}{d\lambda}\frac{dx^{\beta}}{d\lambda},
\ee
where 
\be
	\Gamma^{\mu}_{\alpha\beta} = \frac{1}{2}g^{\mu\nu}\left ( \frac{\partial g_{\alpha\nu}}{\partial x^{\beta}} + \frac{\partial g_{\nu\beta}}{\partial x^{\alpha}} - \frac{\partial g_{\alpha\beta}}{\partial x^{\nu}}\right ) 
\ee
is the metric connection and $\lambda$ is an affine parameter. Using the initial conditions at the image plane of the observer, as defined in Appendix~\ref{app:initials}, we trace photons backwards in time with this equation until it hits the disk (in the case of iron line spectroscopy, see Sec~\ref{sec:xrs}) or satisfies one of the conditions in expression~\ref{eq:condition} (in the case of BH imaging, see Sec~\ref{sec:imaging}).

We now define some quantities related to the accretion disk. For a particle falling freely in a static spacetime with 4-momentum
\begin{equation} \label{eq:fourmom}
p^{\alpha} = \frac {d x^{\alpha}} {d \tau},
\end{equation}
we have (at least) two conserved quantities: (i) the energy $E = -p_{t}$, and (ii) the axial angular momentum $L_{z} = p_{\phi}$ (e.g. \cite{chandbh22}). If we specialize to motion in the equatorial plane $(p^{\theta} = 0)$, appropriate for a thin, unwarped accretion disk, the geodesic equations reduce to
\begin{equation} \label{eq:geodesic}
\frac {1} {2} g_{rr} \left( p^{r} \right)^{2} = - \frac {1} {2} \left[  g_{tt} \left( p^{t} \right)^2  + g_{\phi\phi} \left( p^{\phi} \right)^2 + 1 \right].
\end{equation}
The right-hand side of Eq.~\ref{eq:geodesic} defines the effective potential, $V_{\text{eff}}$. Circular orbits are characterized by the vanishing of the radial momentum and its derivative, i.e. $p^{r} = 0$ and $\dfrac {d p^{r}} {d r}= 0$ \cite{chandbh22}. Particularizing to this case, we have from Eq.~\ref{eq:geodesic}
\begin{equation} \label{eq:veff1}
V_{\text{eff}} = 0 ,
\end{equation}
and
\begin{equation} \label{eq:veff2}
\frac {d} {d r} \left( \frac {V_{\text{eff}}} {g_{rr}} \right) = 0 .
\end{equation}
Solving equations~\ref{eq:veff1} and~\ref{eq:veff2} leads to expressions for the energy $E$ and angular momentum $L_{z}$, which in turn define the Keplerian frequency $\Omega_{\phi}$,
\begin{equation} \label{eq:kepler}
\Omega_{\phi} = -  \frac {g_{tt} L_{z}} {g_{\phi\phi} E},
\end{equation}
the radial epicyclic frequency $\kappa_{r}$,
\begin{equation} \label{eq:radialepi}
\kappa_{r}^2 = - \frac {\partial^2} {\partial r^2} \left[ \frac {V_{\text{eff}}} {g_{rr} \left( p^{t} \right)^{2}} \right]  ,
\end{equation} 
and the angular epicyclic frequency $\Omega_{\theta}$,
\begin{equation} \label{eq:angularepi}
\Omega_{\theta}^2 = -  \frac {\partial^2} {\partial \theta^2} \left[ \frac {V_{\text{eff}}} {g_{\theta\theta} \left( p^{t} \right)^{2}} \right] ,
\end{equation}
where quantities are evaluated in the equatorial plane $\theta = \pi/2$. The expressions for these quantities as functions of $r$, $E$, and $L_{z}$ are long but can be easily evaluated using the metric components in Eq.~\ref{eq:afunct} and~\ref{eq:bfunct}.

Using these quantities, we can get the innermost stable circular orbit and the redshift factor. The former is located at the radial coordinate where 
\be
	\frac{\partial^2V_{\rm eff}}{\partial r^2} = 0 \quad \textrm{ or } \quad \frac{\partial^2V_{\rm eff}}{\partial\theta^2} = 0.
\ee
The latter is defined as
\be
	g = \frac{\nu_o}{\nu_e} = \frac{-u_o^{\alpha}p_{\alpha}}{-u_e^{\beta}p_{\beta}},
\ee
where $\boldsymbol{u_o}$ and $\boldsymbol{u_e}$ are the 4-velocities of the observer at the point of detection and the gas at the point of emission, respectively, and $\boldsymbol{p}$ is the 4-momentum of the photon. Taking the distant observer to be at rest, we get $u_o^{\alpha} = (1,0,0,0)$, and for the gas moving on circular orbits in the accretion disk, we get $u_e^{\beta} = g^{tt}p_t(1,0,0,\Omega_{\phi})$. Plugging these expressions in, the equation for $g$ becomes
\be \label{eq:redshifteq}
	g = \frac {\sqrt{-g_{tt} -g_{\phi \phi} \Omega_{\phi}^2}} {1 + p_{\phi} \Omega_{\phi} / p_{t} }.
\ee
\section{The transfer function formalism\label{app:trf}}
The transfer function serves as an integration kernel that can convert the local spectrum at the point of emission (in our case, the surface of the accretion disk) to the observed spectrum at a detector. It is defined as 
	\be\label{eq-trf}
f(g^*,r_e,\iota) = \frac{1}{\pi r_e} g 
\sqrt{g^* (1 - g^*)} \left| \frac{\partial \left(X,Y\right)}{\partial \left(g^*,r_e\right)} \right| \, .
\ee
where $\left| \frac{\partial \left(X,Y\right)}{\partial \left(g^*,r_e\right)} \right| $ is the Jacobian and $\iota$ is the inclination of the distant observer relative to the normal to the disk (which will differ in general from $\vartheta_e$ due to light bending in the vicinity of the BH). Here, $g^*$ is the normalized redshift factor, defined as
\bead
	g^* = \frac{g - g_{\rm min}}{g_{\rm max} - g_{\rm min}},
\eead
where $g_{\rm max}=g_{\rm max}(r_e,D,\iota)$ and $g_{\rm min}=g_{\rm min}(r_e,D,\iota)$ are the maximum and minimum values of $g$, respectively, and $g, r_e, X, Y$ and $\vartheta_e$ have been defined in Sec.~\ref{subsec:ironline}. For each value of $g^*$ except $0$ and $1$, there are two values of $g$ at a constant $r_e$, resulting in a splitting of a constant radius ring in the accretion disk in two sections. So the transfer function too will have two values at each $g^*$, along each section.

The reformulated flux equation is given as
\begin{widetext}
\bead\label{eq-Fobs2}
	F_o (\nu_o) 
	= &\frac{1}{D^2} \int_{r_{\rm in}}^{r_{\rm out}} \int_0^1 \frac{\pi r_{ e} g^2}{\sqrt{g^* (1 - g^*)}} f^{(1)}(g^*,r_{e},\iota) I_{  e}(\nu_{ e},r_{ e},\vartheta^{(1)}_{ e}) \, dg^* \, dr_{ e} \\
	& + \frac{1}{D^2} \int_{r_{\rm in}}^{r_{\rm out}} \int_0^1 \frac{\pi r_{ e} g^2}{\sqrt{g^* (1 - g^*)}} f^{(2)}(g^*,r_{e},\iota) I_{  e}(\nu_{ e},r_{ e},\vartheta^{(2)}_{ e}) \, dg^* \, dr_{ e},
\eead
\end{widetext}
where $(f^{(1)},\vartheta^{(1)}_e)$ and $(f^{(2)},\vartheta^{(2)}_e)$ are the transfer functions and the emission angles along the two sections, respectively.

Following the \textsc{relxill\_nk} framework~\cite{relxillnk}, the transfer function is calculated at $100$ values of $r_e$ and 20 values of $g^*$ at a specific $\Sigma$ and $\iota$ and stored in a FITS (Flexible Image Transport System) table~\cite{relxillnk}.  
To calculate the flux from the transfer function tables, we interpolate the transfer function in $(r_e,g^*)$~\cite{relxillnk} and then integrate along $r_e$ from the inner to the outer edge of the disk and along $g^*$ from 0 to 1~\cite{speith1995}.  

\bibliography{references}

\end{document}